\documentclass[twocolumn,aps,superscriptaddress]{revtex4}
\usepackage{graphicx}
\usepackage{float}
\usepackage{dcolumn}
\usepackage{bm}
\usepackage{color}
\usepackage{amsmath}
\usepackage{amssymb}
\usepackage{amsfonts}
\usepackage{esint}
\usepackage{times}
\usepackage{xcolor}
\usepackage{phaistos}
\usepackage{braket}
\usepackage{comment}
\usepackage{multirow}

\usepackage{pdfpages}

\usepackage[colorlinks,linkcolor=blue,anchorcolor=blue,urlcolor=blue,urlcolor=blue,citecolor=blue]{hyperref}

\begin{document}

\title{Long-range tunable coupler for modular fluxonium quantum processors}

\author{Peng Zhao}
\email{shangniguo@sina.com}
\affiliation{Quantum Science Center of Guangdong-Hong Kong-Macao Greater Bay Area, Shenzhen 518045, China}
\author{Peng Xu}
\affiliation{Institute of Quantum Information and Technology,
Nanjing University of Posts and Telecommunications, Nanjing, Jiangsu 210003, China}
\author{Zheng-Yuan Xue}
\email{zyxue83@163.com}
\affiliation{Key Laboratory of Atomic and Subatomic Structure and Quantum Control (South China Normal University), Ministry of Education, Guangdong Basic Research Center of Excellence for Structure and Fundamental Interactions of Matter, and School of Physics, South China Normal University, Guangzhou 510006, China}
\affiliation{Guangdong Provincial Key Laboratory of Quantum Engineering and Quantum Materials, Guangdong-Hong Kong Joint Laboratory of Quantum Matter, and Frontier Research Institute for Physics, South China Normal University, Guangzhou 510006, China}
\affiliation{Quantum Science Center of Guangdong-Hong Kong-Macao Greater Bay Area, Shenzhen 518045, China}

\date{\today}

\begin{abstract}

The path toward practical superconducting quantum processors requires the integration
of a large number of high-performance qubits. Modular architectures could offer a way to address the
scaling limitations of monolithic designs by partitioning a large quantum processor into physically
separated modules, or chiplets, linked through long-range interconnects. In this context, although fluxonium
qubits have emerged as a compelling platform for quantum computing due to their
long coherence times and high-fidelity gates, existing coupling schemes remain restricted to qubits
in close proximity on a single chip. This limitation inherently precludes the long-range interconnects essential
for modular integration. In this work, we propose a long-range tunable coupler designed to interconnect
fluxonium qubits separated by more than one centimeter, thereby supporting the realization of modular
fluxonium quantum processors. Under realistic assumptions, the proposed coupler has the potential to achieve inter-module
two-qubit gate performance, specifically sub-100-ns gates with intrinsic errors below $10^{-4}$, comparable to that of
intra-module (intra-chiplet) gates, while enabling modular integration with low
quantum crosstalk, a key requirement for scalable systems. We further discuss the integration of this coupler
into modular fluxonium lattices and explore its feasibility for achieving the higher connectivity and
longer coupling range required for complex quantum error correction codes. This work could contribute to the development of
large-scale fluxonium quantum processors by bridging their demonstrated potential with modular scalability.

\end{abstract}

\maketitle


\section{Introduction}\label{SecI}

The realization of practical superconducting quantum computing ultimately demands the integration of a large
number of high-performance superconducting qubits~\cite{Fowler2012,Gidney2021,Gidney2025}. However, fabricating
and controlling such large-scale devices in monolithic architectures presents formidable scientific and technical
obstacles. As the qubit count increases, wiring complexity grows significantly, leading to severe input-output
bottlenecks~\cite{Frankea2019,Reilly2019,Zhao2024,Mohseni2024} and heat load issues~\cite{Kawabata2026}.
At the same time, physical proximity exacerbates crosstalk between neighboring elements~\cite{Barends2014,Huang2021},
while the stringent fabrication requirements for dense arrays of qubits result in diminished fabrication
and parameter yields~\cite{Mohseni2024,Damme2024}. These interrelated constraints fundamentally limit the scale
achievable on a single chip, motivating the exploration of alternative architectures that can circumvent these
bottlenecks while preserving performance.

Modular architectures are
positioned as a promising paradigm in this direction. By partitioning a
large processor into smaller, physically separated modules (chiplets), the aforementioned challenges
can be localized within each module, while inter-module interconnections maintain
system-wide connectivity~\cite{Gold2021, Zhao2022, Smith2022, Bravyi2022}. In such architectures, the performance of
inter-chiplet interconnections is just as critical as that of intra-chiplet connections in determining overall
system performance. Addressing the scaling challenge through this approach therefore necessitates not only the
construction of high-performance qubit modalities, but also the development of coupling schemes capable of interconnecting
physically separated qubits without compromising system performance and functionality~\cite{Gold2021,Zhao2022,Field2024,Wu2024,Ihssen2025}.

\begin{figure}[tbp]
\begin{center}
\includegraphics[keepaspectratio=true,width=\columnwidth]{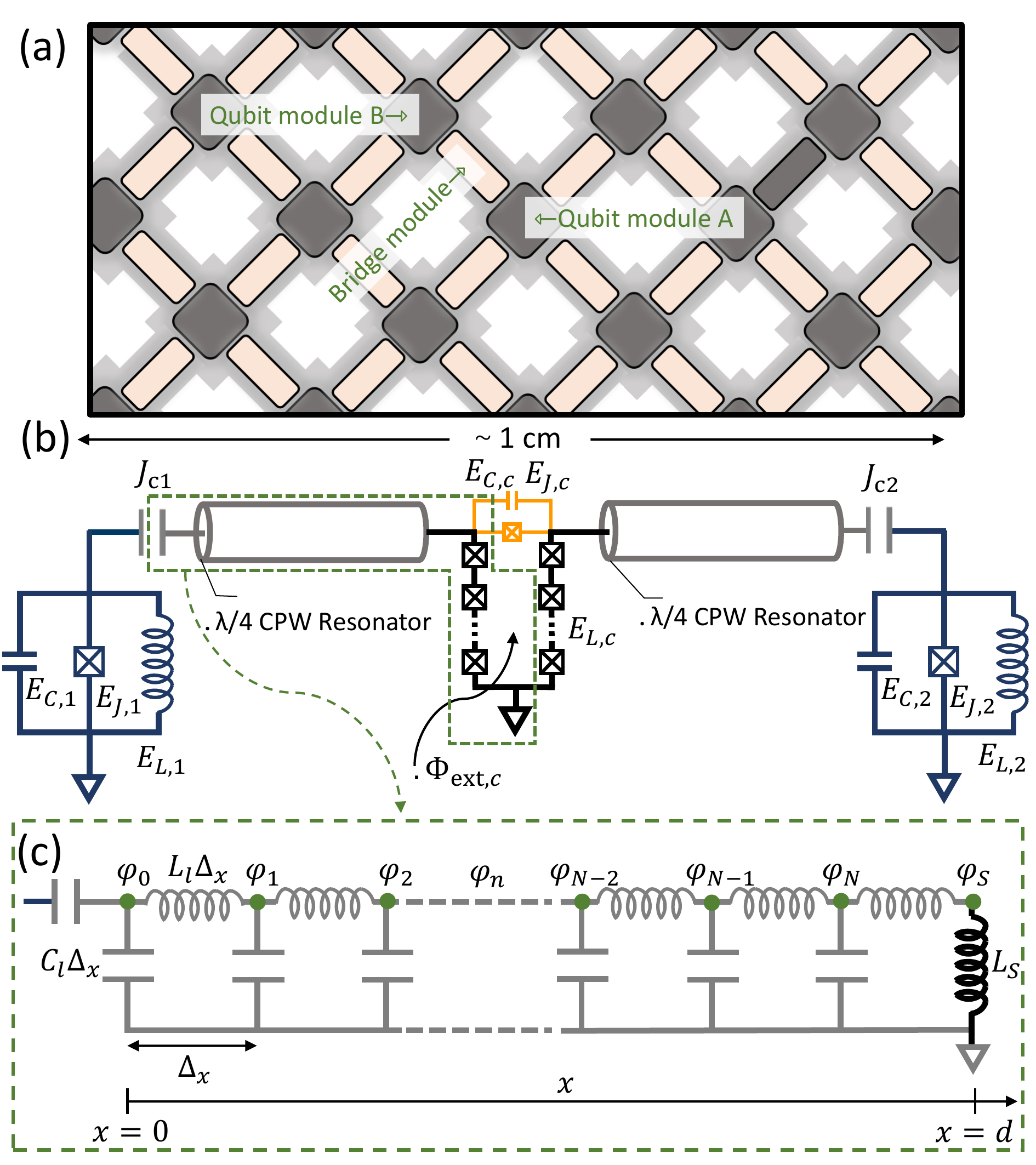}
\end{center}
\caption{A modular fluxonium quantum processor enabled by long-range tunable inter-module couplers.
(a) Schematic illustration of a modular quantum processor architecture. Qubit modules (e.g., modules A and B, each hosting tens or hundreds of qubits or more) are interconnected via long-range couplers hosted on dedicated bridge modules. (b) Schematic of the long-range tunable coupler designed to
interconnect fluxonium qubits separated by one centimeter. The coupler consists of two quarter-wave ($\lambda/4$) superconducting
coplanar waveguide (CPW) resonators coupled via a tunable inductive coupler (rf SQUID coupler). Each $\lambda/4$ CPW resonators is
terminated by a linear inductor $L_{s}$ (a junction array) to ground and can be modeled as a series of identical
lumped LC oscillators with capacitance $C_{l}\Delta_x$ and inductor $L_{l}\Delta_x$, as schematically shown in (c).}
\label{fig1}
\end{figure}

In this context, modular architectures have been explored as a scalable paradigm for superconducting
circuits, with transmon qubits~\cite{Koch2007} constituting the foundation of prevailing
explorations~\cite{Gold2021, Zhao2022, Smith2022, Bravyi2022,Field2024,Wu2024,Ihssen2025}. A specific
example is a recently proposed transmon-based design~\cite{Zhao2022}, in which qubit modules (each hosting tens or hundreds 
of qubits or more) are spatially separated by centimeters
and interconnected via long-range tunable couplers hosted on dedicated bridge modules, as shown in Fig.~\ref{fig1}(a) (a similar design
has also been demonstrated in a small-scale implementation in Ref.~\cite{Ihssen2025}). This approach could effectively
decouple fabrication across modules, allowing individual modules to be tested and selected prior to integration~\cite{Yao2026}.
Thus, such a strategy could potentially not only improve overall qubit yield but also enhance parameter yield by enabling the assembly of modules
with consistently matched device specifications, including frequencies and coupling strengths. Moreover, beyond enabling
physical separation, long-range interconnects offer
an additional advantage. By providing space between qubit modules [e.g., white regions in Fig.~\ref{fig1}(a)], they could facilitate
the compact integration of Cryo-CMOS
electronics~\cite{Patra2018} or superconducting digital logic, e.g., Single Flux Quantum (SFQ)~\cite{Likharev1991} and
Adiabatic Quantum Flux Parametron (AQFP)~\cite{Takeuchi2013}, for qubit control~\cite{Patra2018,McDermott2018,Leonard2019} within the interstitial
regions. The compact integration of control electronics enabled by this approach could create new opportunities for more efficient
control architectures~\cite{Acharya2023, Bernhardt2025, Liu2026, Takeuchi2024}. Such placement holds substantial promise for alleviating control-line routing complexity and resolving the
associated wiring bottlenecks~\cite{Frankea2019}. More importantly, the tunable interaction between
inter-chiplet qubit pairs, enabled by long-range couplers, is essential for suppressing quantum crosstalk between modules
and ensuring high-fidelity intra- and inter-chiplet gate operations~\cite{Zhao2022,Field2024,Xu2026}.

Fluxonium qubits~\cite{Manucharyan2009} have emerged as a compelling alternative to transmons for
scalable quantum processors, owing to their long coherence times~\cite{Somoroff2023} and high-fidelity gate operations~\cite{Somoroff2023,Rower2024,Ding2023,Zhang2024,Lin2025}. However, they will inevitably encounter the same
scaling constraints that limit monolithic architectures. Developing a modular architecture therefore presents an
appealing pathway toward large-scale fluxonium quantum processors. Despite this promise, a critical challenge persists, i.e., the coupling
schemes demonstrated or explored to date, which are primarily based on direct inductive~\cite{Lin2025,Nesterov2018,Ma2024} or capacitive~\cite{Nesterov2018,Ficheux2021,Bao2022,Chen2022,Nesterov2022,Dogan2023} interactions, or indirect coupler-mediated interactions~\cite{Ding2023,Zhang2024,Moskalenko2021,Moskalenko2022,Weiss2022,Simakov2023,Rosenfeld2024,Xiong2025,Zhao2025,Zhao2026}, are
inherently limited to qubits in close proximity on a single chip. More importantly, for fluxoniums coupled with always-on interactions, whether
directly or indirectly, significant quantum crosstalk arises~\cite{Zhao2025,Zhao2026,Zwanenburg2026}, which can substantially degrade system performance. Although
such issue could be partially mitigated through strategies such as careful frequency allocation~\cite{Nesterov2018,Kugut2025,Moreno2025}, this
approach imposes additional burdens on the fabrication process, analogous to the challenges encountered in scaling qubit architectures
based on fixed-coupled, fixed-frequency transmon qubits~\cite{Hertzberg2020}.

The above limitations of existing fluxonium systems pose a fundamental obstacle to
develop modular integration, as realizing modular architectures demands long-range, low-crosstalk interconnects that remain largely
unexplored for fluxonium qubits. Moreover, recent efforts toward implementing
quantum low-density parity-check (qLDPC) codes further reinforce the case for long-range coupling~\cite{Breuckmann2021,Bravyi2024,Shaw2025}.
Thus, in this work, to advance from the demonstrated potential of fluxonium qubits toward the requirements of modular scalability, we
propose and theoretically analyze a long-range tunable coupler designed to interconnect fluxonium qubits. This coupler is
based on two quarter-wave superconducting coplanar waveguide (CPW) resonators coupled via a tunable inductive
coupler. Such a design can enable tunable coupling between fluxonium qubits separated by distances at the centimeter scale.

Under physically realistic assumptions, we demonstrate that this coupler design can realize high-contrast
two-fluxonium interactions, both for suppressing quantum crosstalk and for enabling fast, high-fidelity
two-qubit gates. Specifically, sub-100-ns CZ gates for fluxonium qubits separated by more than one centimeter
can be achieved with intrinsic errors below $10^{-4}$. Building on this, we further show the integration of this coupler
into modular fluxonium lattices with degree-four connectivity and demonstrate its potential to achieve the
higher-degree connectivity required for complex quantum error correction codes. Finally, we suggest that
when such a coupler design is combined with a multimode cavity QED architecture~\cite{McKay2015}, even long-range and high-degree
connectivity could potentially be realized with low quantum crosstalk.

By extending the modular paradigm to the fluxonium platform and providing a viable pathway for long-range, low-crosstalk
coupling, this work could offer a stepping stone toward large-scale, modular quantum processors based on fluxonium
qubits and could also offer insights that may help address the wiring, crosstalk, and yield challenges that impede
monolithic approaches. It could also facilitate the realization of complex quantum error correction codes that require
long-range, high-degree connectivity. Furthermore, beyond fluxonium qubits, the mechanisms of the coupler design revealed
in this work are also applicable to other types of superconducting qubits, including the transmon qubit.

This paper is organized as follows. In Sec.~\ref{SecII}, we introduce the long-range tunable coupler
for fluxonium qubits and analyze the tunable interaction between two fluxoniums connected via this
design. In Sec.~\ref{SecIII}, we numerically study microwave-activated CZ gates
in the two-fluxonium system presented in Sec.~\ref{SecII}, analyze the dominant intrinsic error sources, and
extend our analysis to gates performed with longer couplers. In Sec.~\ref{SecIV}, we discuss the challenges and
opportunities of integrating the long-range tunable coupler into fluxonium qubit lattices with degree-four connectivity
within a modular architecture. We also assess the feasibility of leveraging this coupler to achieve the
long-range, higher-degree connectivity necessary for implementing complex quantum error correction codes.
We further outline near-term directions for extending the coupler length and enhancing connectivity.
Finally, in Sec.~\ref{SecV}, we present the conclusions of our work.

\section{circuit model and system Hamiltonian}\label{SecII}

Here, we begin by presenting the design and operating principles of the long-range tunable
coupler (LTC),  focusing on its internal tunability. Subsequently, we discuss its ability to mediate tunable interactions
between qubits over extended distances and consider the configuration in which two fluxonium qubits are coupled through
this coupler and analyze the resulting tunable coupling between them.

\subsection{Long-range tunable coupler}\label{SecIIA}

A schematic of the LTC circuit considered here is shown in Fig.~\ref{fig1}(b). The coupler
consists of two quarter-wave $\lambda/4$ superconducting coplanar waveguide (CPW) resonators.
Each $\lambda/4$ CPW resonator is terminated to ground by a linear inductor $L_s$, which could
be implemented as a Josephson junction array~\cite{Masluk2012,Bell2012}. Similar to the gmon
coupler design used for coupling transmon qubits~\cite{Chen2014,Geller2015}, the two resonators
here are inductively coupled via a coupling junction. This junction, together with the two linear
inductors, forms an rf-SQUID coupler embedded within the LTC. Accordingly, tuning the flux through
the rf-SQUID loop modulates the mutual inductance between the two CPW resonators, enabling their
coupling strength to be dynamically turned on and off. This flux tunability is essential for both
suppressing quantum crosstalk between fluxonium qubits and enabling fast two-qubit gates, as will
be discussed in the following section.

In this subsection, before delving into the full system description of the LTC, we first provide
a more systematic analysis of the $\lambda/4$ CPW resonator terminated by a linear
inductor, following Ref.~\cite{Wallquist2006}. Subsequently, we turn to derive the full system
Hamiltonian of the LTC coupler.

\subsubsection{$\lambda/4$ CPW resonator terminated with a linear inductor}\label{SecIIA1}

By partitioning the CPW structure of the resonator into $N$ identical sections
of total length $d = N \Delta_x$, the CPW part is modeled as a series of $N$ identical lumped LC
oscillators with capacitance $C_{l}\Delta_x$ and inductance $L_{l}\Delta_x$ [see Fig.~\ref{fig1}(c)].
Here, $C_{l}$ and $L_{l}$ denote the capacitance and inductance per unit length of the CPW structure, respectively.
Considering the phase $\varphi_{n}$ across the $n$-th capacitor and its time derivative $\dot{\varphi}_{n}$, the
Lagrangian of the CPW resonator is given by
\begin{equation}
\begin{aligned}\label{eq1}
\mathcal{L}_{c}=\mathcal{L}_{TL}+\mathcal{L}_{S},
\end{aligned}
\end{equation}
where
\begin{equation}
\begin{aligned}\label{eq2}
\mathcal{L}_{TL}=&\sum_{n=0}^{N-1}\left\{\left(\frac{\hbar}{2e}\right)^2
\left[\frac{C_{l}\Delta_{x}\dot{\varphi}_{n}^2}{2}-\frac{(\varphi_{n+1}-\varphi_{n})^2}{2L_{l}\Delta_{x}}\right]\right\}
\\&+\left(\frac{\hbar}{2e}\right)^2\left[\frac{C_{l}\Delta_{x}\dot{\varphi}_{N}^2}{2}
-\frac{(\varphi_{S}-\varphi_{N})^2}{2L_{l}\Delta_{x}}\right]
\end{aligned}
\end{equation}
corresponds to the CPW part of the resonator and
\begin{equation}
\begin{aligned}\label{eq3}
\mathcal{L}_{S}=-\left(\frac{\hbar}{2e}\right)^2\frac{\varphi_{S}^2}{2L_{S}}
\end{aligned}
\end{equation}
describes the linear inductor part, with $\varphi_{S}$ denoting the phase across the linear inductor.

Taking the continuum limit $\Delta_x\rightarrow 0$, the discrete phase variable $\varphi_n$ becomes the continuous
variable $\varphi(x)$. Away from the terminated section of the resonator, the corresponding Euler-Lagrange equation
for the phase dynamics is
\begin{equation}
\begin{aligned}\label{eq4}
\nu^2\frac{\partial^2\varphi(x,t)}{\partial x^2}-\frac{\partial^2\varphi(x,t)}{\partial t^2}=0,
\end{aligned}
\end{equation}
where $\nu=1/\sqrt{L_{l}C_{l}}$ denotes the wave velocity of the CPW structure. Given the open
boundary condition $\partial\varphi(0,t)/\partial x=0$, the solutions to the above equation take the
form $\varphi_{k}(x,t)=\phi\sin(k\nu t)\cos(kx)=\varphi_{k}(t)\cos(kx)$. Here, $k$ is the wave vector of an eigenmode
of the resonator, and any solution of the wave equation can be written as a linear combination of
these eigenmode solutions. Further considering the boundary condition at the terminated linear inductor,
\begin{equation}
\begin{aligned}\label{eq5}
\varphi_{k}(d,t)=\varphi_{S}(t),
\end{aligned}
\end{equation}
leads to a transcendental equation that constrains these eigenmode solutions:
\begin{equation}
\begin{aligned}\label{eq6}
kd\tan(kd)=\frac{dL_{l}}{L_{S}}.
\end{aligned}
\end{equation}

The solution to Eq.~(\ref{eq6}) defines an infinite set of eigenmodes within the CPW resonator. Each eigenmode
is characterized by the wave vector $k$, which gives rise to the eigenmode frequency $\omega_k = k\nu = k/\sqrt{L_l C_l}$.
Accordingly, the CPW resonator can now be described as a collection of (decoupled) lumped LC oscillators, each
corresponding to one eigenmode defined by Eq.~(\ref{eq6}). The Lagrangian of the CPW resonator can thus be
expressed as a sum of the Lagrangians $\mathcal{L}_k$ of the individual eigenmodes (each treated as an LC
oscillator), given by~\cite{Wallquist2006}
\begin{equation}
\begin{aligned}\label{eq7}
\mathcal{L}_{k}=\left(\frac{\hbar}{2e}\right)^2\left[\frac{C_{k}\dot{\varphi}_{k}(t)^2}{2}-\frac{\varphi_{k}^2(t)}{2L_{k}}\right]
\end{aligned}
\end{equation}
with
\begin{equation}
\begin{aligned}\label{eq8}
&C_{k}=\frac{dC_{l}}{2}\left[1+\frac{\sin(2kd)}{2kd}\right],
\\&L_{k}=\frac{2dL_{l}}{(kd)^2}\left[1+\frac{\sin(2kd)}{2kd}\right]^{-1}
\end{aligned}
\end{equation}
representing the effective capacitance and inductance of the LC oscillator for the eigenmode with
wave vector $k$.

Despite the multimode nature of the $\lambda/4$ CPW resonator, we hereafter mainly focus on its fundamental mode.
This is a reasonable assumption when choosing appropriate resonator parameters, as the frequencies of the
higher-harmonic modes can be far beyond those relevant to our coupled system. In this case, it suffices to retain
only the fundamental mode. To be more specific, we adopt the CPW structure parameters summarized in
Table~\ref{tab:CPW_parameters}. This allows the construction of a specific LTC with a length of one
centimeter, thus enabling the connection of fluxonium qubits separated by the same distance.
In the following discussion, unless stated otherwise, the numerical results presented below are obtained using
these parameters. Moreover, we note that the design and verification of the LTC's CPW structure
are carried out using Wcalc~\cite{Wcalc}, a tool dedicated to electromagnetic analysis.

\begin{table}[!htb]
\caption{\label{tab:CPW_parameters}
Characterization of the CPW structure. The parameter values listed in the table include the total length $d$, the
capacitance $C_{l}$ and inductance $L_{l}$ per unit length, and the characteristic impedance $Z_{0}$ of the
CPW structure.}
\begin{ruledtabular}
\begin{tabular}{cccc}
$ d$ (cm)& $L_l$ (nH/cm)& $C_l$ (fF/cm) & $Z_{0}$ ($\Omega$) \\\hline
0.50 & 4.37 & 1586.42 & 52.47 \\
\end{tabular}
\end{ruledtabular}
\end{table}

\begin{figure*}[tbp]
\begin{center}
\includegraphics[width=16cm,height=7cm]{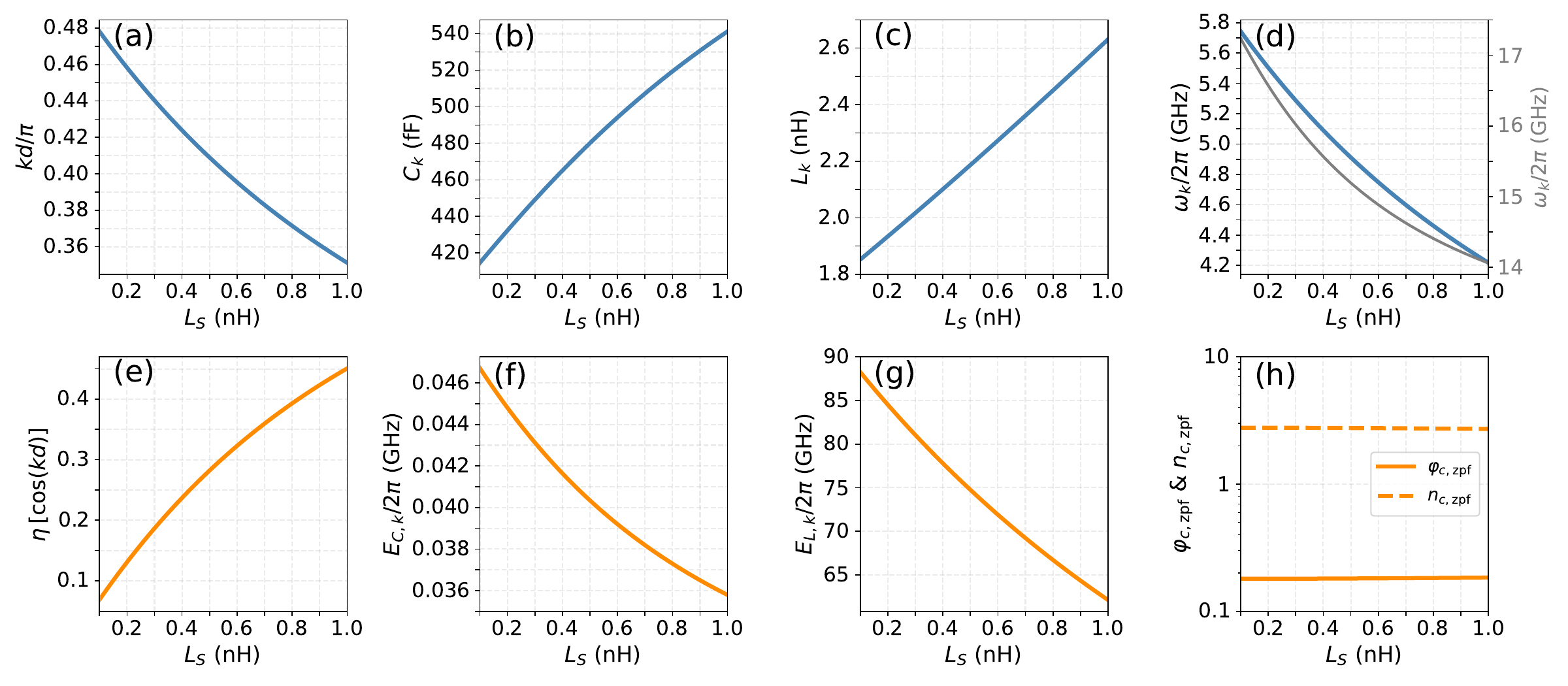}
\end{center}
\caption{Parameters of the fundamental mode of the CPW resonator terminated with a linear inductor.
(a) The wave vector $k$ of the fundamental mode, (b) the effective capacitance, (c) the effective
inductance, and (d) the fundamental mode frequency as functions of the inductance $L_{S}$. For easy
reference, the frequency of the first excited mode is also shown in (d). (e) The profile of the phase
field of the fundamental mode at the end of the CPW structure [$x=d$, see Fig.~\ref{fig1}(c)] versus
$L_{s}$. (f) The effective charge energy and (g) the effective inductive energy of the fundamental
mode versus $L_{s}$. (h) The zero-point fluctuations of the phase and the charge number for the
fundamental mode versus $L_{s}$.}
\label{fig2}
\end{figure*}

Given these CPW parameters and according to Eq.~(\ref{eq5}), Fig.~\ref{fig2}(a) shows the wave vector of the
fundamental mode of the resonator as a function of the terminating inductor $L_{S}$. As expected, when $L_{S}$
decreases, the product $kd$ approaches $\pi/2$, corresponding to a standard $\lambda/4$ CPW resonator.
According to Eq.~(\ref{eq7}), the dependence of the effective inductance and capacitance of the fundamental
mode on the terminating inductor is also displayed, as shown in Figs.~\ref{fig2}(b) and~\ref{fig2}(c).

Furthermore, Fig.~\ref{fig2}(d) shows the frequencies of both the fundamental and first harmonic modes of the
resonator for different terminating inductors. As mentioned earlier, the higher-mode frequency is far
beyond the frequency range of interest in this work. Therefore, in the following discussion, we mainly
focus on the fundamental mode of the resonator.

\subsubsection{Strongly coupled two-mode system with high on-off ratio}\label{SecIIA2}

Here, we turn to derive the full system Hamiltonian of the LTC coupler. Focusing on the fundamental modes of the two CPW
resonators within the LTC and recalling that $\varphi_{S}(t)=\cos(kd)\varphi_{k}(t)\equiv\eta\varphi_{k}(t)$
[see Fig.~\ref{fig2}(e) for the dependence of the phase field profile at $x = d$, i.e., $\eta$, on $L_{S}$], the full Lagrangian
of the LTC is given by
\begin{equation}
\begin{aligned}\label{eq9}
&\mathcal{L}_{\rm LTC}=\mathcal{L}_{c1}+\mathcal{L}_{c2}+\mathcal{L}_{CJJ}
\\&\mathcal{L}_{CJJ}=\left(\frac{\hbar}{2e}\right)^2\frac{C_{J}}{2}\left[\eta_{1}\dot{\varphi}_{c1}(t)-\eta_{2}\dot{\varphi}_{c2}(t)\right]^2
\\&\quad\quad\quad+E_{J,c}\cos(\eta_{1}\varphi_{c1}(t)-\eta_{2}\varphi_{c2}(t)-\varphi_{{\rm ext},c}).
\end{aligned}
\end{equation}
In the above expression, $\mathcal{L}_{c1(c2)}$ denotes the Lagrangian of each CPW resonator, which takes the form
shown in Eq.~(\ref{eq7}), and $\mathcal{L}_{CJJ}$ describes the resonator-resonator coupling mediated by
the coupling junction [see Fig.~\ref{fig1}(a)]. It is worth noting that, due to the distributed nature of
the CPW structure, the phase profile of the fundamental mode is spatially dependent and takes the form $\cos(kd)$.
This gives rise to the factor $\eta_j$ appearing in the expression for $\mathcal{L}_{CJJ}$. Here, we have also
introduced the external flux bias $\varphi_{\text{ext}} = 2\pi \Phi_{\text{ext}} / \Phi_0$, where $\Phi_0$ denotes
the flux quantum. Note also that while the coupling junction mediates both inductive ($E_{J,c}$) and
capacitive ($C_J$) interactions, the inter-resonator coupling is dominated by the inductive
interaction, as will be discussed later.

Given Eqs.~(\ref{eq9}) and~(\ref{eq7}), the LTC coupler can be described by the following Hamiltonian (similar in
form to that of the ultrastrong inter-resonator coupler~\cite{Mukai2019,Miyanaga2021} and the double-transmon coupler~\cite{Goto2022,Campbell2023})
\begin{equation}
\begin{aligned}\label{eq10}
H_{\rm LTC}=& \sum_{j=1,2} [4 E_{C,cj} n_{cj}^2 + \frac{E_{L,cj}}{2}\varphi_{cj}^2]
\\ &+ 4E_{C,c}\eta_{1}\eta_{2} n_{c1} n_{c2}
\\ &-E_{J,c}\cos(\eta_{1}\varphi_{c1}-\eta_{2}\varphi_{c2}-\varphi_{{\rm ext},c}),
\end{aligned}
\end{equation}
where the subscripts ${j=1, 2}$ index the two resonator modes, $E_{C,cj}=e^2/(2C_{j})$ and $E_{L,cj}=(\hbar/2e)^2/L_{j}$ denote the
charge and inductive energy, respectively [see Figs.~\ref{fig2}(f,g) for their dependence on the $L_{S}$]. Here, the terms
in the first line describe the two resonator modes, the second line describes the inter-resonator capacitive coupling, mediated e.g., by the intrinsic junction capacitance, with strength $E_{C,c} = e^2 C_J / (C_1 C_2)$, and the third line describes the inter-resonator nonlinear
inductive coupling mediated by the coupling junction.

Note that, as indicated in Ref.~\cite{Bourassa2009}, in the derivation of the above Hamiltonian, we assume
that the coupling junction does not significantly perturb the CPW resonators. This ensures that the mode
approximation for $\varphi$ derived in the previous subsection remains valid even in the presence of the
coupling junction. This approximation is justified by considering the following: (i) the coupling capacitance
$C_J$ (typically a few fF) is generally much smaller than the effective capacitance of the
resonator [see Fig.~\ref{fig2}(b)]; (ii) the inductance of the coupling junction can be made larger than the
terminating linear inductor $L_S$, so that most of the resonator current flows through the terminating
inductor to ground.

Generally, when the LTC circuit is used as a coupler, it is assumed to be in its ground state. Here, we thus
concern ourselves only with the lower-lying energy states of the coupler. Accordingly, by expanding the potential
energy of the LTC,
\begin{equation}
\begin{aligned}\label{eq11}
U_{\rm LTC}=&\frac{E_{L,c1}}{2}\varphi_{c1}^2+\frac{E_{L,c2}}{2}\varphi_{c2}^2
\\&-E_{J}\cos(\eta_{1}\varphi_{c1}-\eta_{2}\varphi_{c2}-\varphi_{{\rm ext},c}),
\end{aligned}
\end{equation}
around its minimum (equilibrium) position ($\bar{\varphi}_{c1},\,\bar{\varphi}_{c2}$) to second
order~\cite{Geller2015,Miyanaga2021,Goto2022}, we obtain the linear approximation of the
potential energy:
\begin{equation}
\begin{aligned}\label{eq12}
\hat U_{\rm LTC}\approx&\sum_{j=1,2}\left[\frac{E_{L,cj}+E_{J,c}\eta_{j}^{2}\cos(\bar{x}-\varphi_{{\rm ext},c})}{2}\hat\varphi_{cj}^{2}\right]
\\&-E_{J,c}\eta_{1}\eta_{2}\cos(\bar{x}-\varphi_{{\rm ext},c})\hat\varphi_{c1}\hat\varphi_{c2},
\end{aligned}
\end{equation}
with
\begin{equation}
\begin{aligned}\label{eq13}
\bar{x}\equiv\eta_{1}\bar{\varphi}_{c1}-\eta_{2}\bar{\varphi}_{c2}.
\end{aligned}
\end{equation}
Here, the minimum position ($\bar{\varphi}_{c1},\,\bar{\varphi}_{c2}$) is defined by the following two relations
\begin{equation}
\begin{aligned}\label{eq14}
&\bar{\varphi}_{c1}+\eta_{1}\frac{E_{J,c}}{E_{L,c1}}\sin(\eta_{1}\bar{\varphi}_{c1}-\eta_{2}\bar{\varphi}_{c2}-\varphi_{{\rm ext},c})=0,
\\&\bar{\varphi}_{c2}-\eta_{2}\frac{E_{J,c}}{E_{L,c2}}\sin(\eta_{1}\bar{\varphi}_{c1}-\eta_{2}\bar{\varphi}_{c2}-\varphi_{{\rm ext},c})=0.
\end{aligned}
\end{equation}

Incorporating Eq.~(\ref{eq12}), the system Hamiltonian of the LTC can be expressed as
\begin{equation}
\begin{aligned}\label{eq15}
\hat{H}_{\rm LTC}\approx & \sum_{j=1,2} \left[4 E_{C,cj} \hat{n}^2_{cj} + \frac{\tilde{E}_{L,cj}(\varphi_{{\rm ext},c})}{2}\hat{\varphi}_{cj}^2\right]
\\&+4E_{C,c}\eta_{1}\eta_{2} \hat{n}_{c1} \hat{n}_{c2}
\\&-E_{J,c}\eta_{1}\eta_{2}\cos(\bar{x}-\varphi_{{\rm ext},c})\hat\varphi_{c1}\hat\varphi_{c2}.
\end{aligned}
\end{equation}
Here, $\tilde{E}_{L,cj}(\varphi_{{\rm ext},c})\equiv E_{L,cj}+E_{J,c}\eta_{j}^{2}\cos(\bar{x}-\varphi_{{\rm ext},c})$
denotes the inductive energy of the resonator, including the contribution from the nonlinear coupling mediated by
the coupling junction. The second term in the Hamiltonian describes a static capacitive coupling between the two
resonators. The third term corresponds to an inductive inter-resonator coupling, which can be dynamically turned
on and off via the flux bias $\varphi_{{\rm ext},c}$.

To explicitly examine the strengths of both the inductive and capacitive inter-resonator couplings within the LTC, we introduce
the creation and annihilation operators for each resonator mode:
\begin{equation}
\begin{aligned}\label{eq16}
&\hat \varphi_{cj} = \phi_{cj,{\rm zpf}}(\hat a_{cj}^{\dag}+\hat a_{cj}),
\quad \hat n_{cj} = i n_{cj,{\rm zpf}}(\hat a_{cj}^{\dag}-\hat a_{cj}),
\end{aligned}
\end{equation}
where $\phi_{c,\rm{zpf}}$ and $n_{c,\rm{zpf}}$ denote the zero-point fluctuations of the phase and charge
number operators, respectively [see Fig.~\ref{fig2}(h) for their dependence on the terminating inductor $L_S$].
These are given by
\begin{equation}
\begin{aligned}\label{eq17}
&\varphi_{cj,{\rm zpf}} =\frac{1}{\sqrt{2}}\left[\frac{8E_{C,cj}}{E_{L,cj}}\right]^{\frac{1}{4}},
n_{cj,{\rm zpf}}= \frac{1}{\sqrt{2}}\left[\frac{E_{L,cj}}{8E_{C,cj}}\right]^{\frac{1}{4}}.
\end{aligned}
\end{equation}
Consequently, the strengths of the static capacitive coupling and the inductive coupling are given by
\begin{equation}
\begin{aligned}\label{eq18}
&g_{c,cap}=-4E_{C,c}\eta_{1}\eta_{2}n_{c1,{\rm zpf}}n_{c2,{\rm zpf}},
\\&g_{c,induc}=-E_{J,c}\eta_{1}\eta_{2}\cos(\bar{x}-\varphi_{{\rm ext},c})\varphi_{c1,{\rm zpf}}\varphi_{c2,{\rm zpf}}.
\end{aligned}
\end{equation}

\begin{figure}[tbp]
\begin{center}
\includegraphics[keepaspectratio=true,width=\columnwidth]{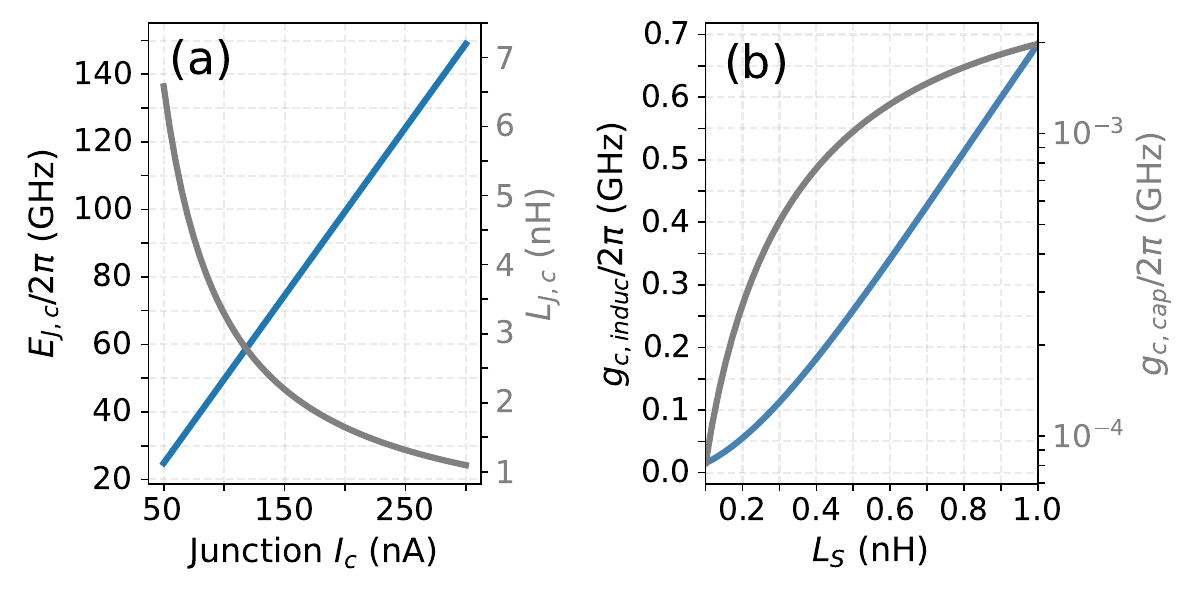}
\end{center}
\caption{Inter-resonator coupling strength mediated by the coupling junction. (a) The junction energy $E_{J,c}$
and the corresponding junction inductance $L_{J,c}$ plotted as functions of the Josephson junction's critical
current. (b) The maximum inductive inter-resonator coupling and the static capacitive inter-resonator coupling
as functions of the terminated inductance $L_{S}$, assuming a junction critical current of $200\,\rm nA$ and an
intrinsic junction capacitance of $2.5\,\rm fF$.}
\label{fig3}
\end{figure}

Based on the above analysis and the CPW parameters in Table~\ref{tab:CPW_parameters}, we now quantitatively characterize
both the inductive and capacitive inter-resonator couplings in the LTC. For simplicity and clarity, we hereafter assume
that the two CPW resonators have identical parameters (see Table~\ref{tab:CPW_parameters}). Figure~\ref{fig3}(a) shows
$E_{J,c}$ and the associated junction inductance $L_{J,c}$ as functions of the junction critical current $I_c$.
In the present work, we take the junction critical current to be $200\,\rm nA$, which corresponds to a junction inductance
of $1.5\,\rm nH$ and a junction energy of $99.336\,\rm GHz$ (see also Table~\ref{tab:fluxonium_parameters}).

Accordingly, assuming an intrinsic junction capacitance of $2.5\,\rm fF$, Figure~\ref{fig3}(b) shows the magnitude
of the static capacitive coupling and the maximum magnitude of the inductive coupling as functions of the
terminating inductor inductance $L_S$. As mentioned earlier, although both capacitive and inductive couplings
arise from the coupling junction, the inductive coupling is the dominant one. Specifically, for $L_S = 0.5\,\rm nH$, the
magnitude of $g_{c,cap}$ is approximately $1\,\rm MHz$, while the maximum magnitude of $g_{c,induc}$ reaches
about $250\,\rm MHz$. This is because the effective capacitance of the resonator is far larger than the coupling
capacitance, thereby heavily suppressing the capacitive coupling energy $E_{C,c}$. This observation
holds even when the junction capacitance is increased from $2.5\,\rm fF$ to $10\,\rm fF$ (a point that will be further
discussed in Fig.~\ref{fig6}(b)).

Since the inductive coupling is the dominant contribution to the inter-resonator coupling, we provide
further details on its tunable control. As indicated by Eq.~(\ref{eq18}), the flux biases for achieving
the maximum magnitude of $g_{c,induc}$ and for turning off the coupling should satisfy the following
relations (where $m$ denotes an integer):
\begin{equation}
\begin{aligned}\label{eq19}
&{\rm ON_{max}:}\bar{x}-\varphi_{{\rm ext},c}=m\pi,
\\&{\rm OFF:}\bar{x}-\varphi_{{\rm ext},c}=\frac{\pi}{2}+m\pi.
\end{aligned}
\end{equation}
Subtracting the two equations in Eq.~(\ref{eq14}), we obtain the following relation for $\bar{x}$ at the equilibrium position
\begin{equation}
\begin{aligned}\label{eq20}
\bar{x}=-\left(\sum_{j=1,2}\eta_{j}^{2}\frac{E_{J,c}}{E_{L,cj}}\right)\sin\left(\bar{x}-\varphi_{{\rm ext},c}\right)
\end{aligned}
\end{equation}
Incorporating Eqs.~(\ref{eq19}) and~(\ref{eq20}), we obtain the following expression for the flux bias that achieves the
maximum coupling magnitude:
\begin{equation}
\begin{aligned}\label{eq21}
\varphi_{{\rm ext},c}\Big|_{\rm ON_{\rm max}}=m\pi
\end{aligned}
\end{equation}
and the flux bias for turning off the coupling (zero-coupling point):
\begin{equation}
\begin{aligned}\label{eq22}
\varphi_{{\rm ext},c}\Big|_{\rm OFF}=&-\left(\sum_{j=1,2}\eta_{j}^{2}\frac{E_{J,c}}{E_{L,cj}}\right)\sin\left(\frac{\pi}{2}+m\pi\right)
\\&-\left(\frac{\pi}{2}+m\pi\right).
\end{aligned}
\end{equation}

\subsection{LTC-mediated tunable fluxonium interactions}\label{SecIIB}

In this subsection, we first provide an analytical analysis of the two-fluxonium interaction mediated by the LTC proposed
in the preceding section. Beyond the enabled long-distance range, we focus on a prominent feature of this coupling
architecture, namely a tunable coupler with an intrinsic coupling-nulling condition that is independent of
the coupled fluxonium qubits. Subsequently, using the full system parameters listed in
Table~\ref{tab:fluxonium_parameters} (the corresponding CPW parameters are already given in
Table~\ref{tab:CPW_parameters}), we consider a specific example and numerically demonstrate that
high-contrast two-fluxonium interactions can be achieved, with the coupling features agreeing well
with the analytical analysis.

\subsubsection{High-contrast two-fluxonium interactions with an intrinsic nulling condition}\label{SecIIB1}

As shown in Fig.~\ref{fig1}(b), we consider two fluxonium qubits $Q_1$ and $Q_2$ coupled via the LTC proposed in the
preceding section. The coupled two-fluxonium system can be described by the following Hamiltonian (hereafter $\hbar = 1$):
\begin{equation}
\begin{aligned}\label{eq23}
\hat{H}= &\sum_{j=1,2} [4 E_{C,j} \hat{n}^2_j+\frac{1}{2}E_{L,j}(\hat\varphi_j - \varphi_{\text{ext},j})^2-E_{J,j}\cos\hat\varphi_j]
\\& + J_{c1} \hat{n}_1 \hat{n}_{c1} + J_{c2} \hat{n}_2 \hat{n}_{c2} +\hat{H}_{LTC},
\end{aligned}
\end{equation}
where $E_C$, $E_J$, and $E_L$ represent the charging, Josephson, inductive energy of the fluxonium qubits, respectively,
$\hat{n}$ and $\hat\varphi$ denote the charge number and the phase operator, and the subscripts~${j=1, 2}$ index the two fluxonium qubits.
Here, $J_{cj}$ represents the strength of the interaction between the $j$-th fluxonium qubit. The Hamiltonian of the
LTC, $\hat{H}_{\text{LTC}}$, is given in Eq.~(\ref{eq10}).

In Eq.~(\ref{eq23}), unless stated otherwise, we assume that each fluxonium qubit is biased at its half-flux-quantum sweet
spot, i.e., $\varphi_{\text{ext},j}/(2\pi) = 0.5$. In addition, we note that hereafter the lowest four fluxonium states are denoted
as ${|0\rangle, |1\rangle, |2\rangle, |3\rangle}$, and the single qubit (computational) subspace refers to the space
spanned by $\{|0\rangle, |1\rangle\}$. The full system state is denoted as $|Q_{1} Q_{2}, \text{LTC}\rangle$, and is
reduced to $|Q_{1} Q_{2}\rangle \equiv |Q_{1} Q_{2}, 0\rangle$ when confined to the two-fluxonium subspace.
Accordingly, state transitions within the qubit subspace are referred to as qubit transitions, such
as $|0\rangle \leftrightarrow |1\rangle$, while transitions between qubit states and non-computational
states are referred to as plasmon transitions, such as $|1\rangle \leftrightarrow |2\rangle$
and $|0\rangle \leftrightarrow |3\rangle$.

Before delving into the details of the coupled system, we note that the fluxonium qubit possesses a
rich, strongly anharmonic spectrum and allows state transitions with frequencies ranging from several
tens of MHz (for qubit transitions) to several GHz (for plasmon transitions). More importantly, the
electric transition dipole for the qubit transition is rather weak, typically about an order of
magnitude smaller than that of transmons, whereas the transition dipole of the plasmon transitions
is comparable to that of transmons.

Here, for general purposes and clarity, we thus first focus on a specific
fluxonium transition $|k\rangle \leftrightarrow |l\rangle$. By inserting Eqs.~(\ref{eq15})-(\ref{eq18}) into
the system Hamiltonian in Eq.~(\ref{eq23}) and applying the rotating-wave approximation (RWA) to the
fluxonium-coupler coupling, we obtain
\begin{equation}
\begin{aligned}\label{eq24}
&\hat{H}_{p}=\sum_{j=1,2}[\omega_{p,j}\hat p_{j}^{\dag}\hat p_{j}]+\hat{H}_{LTC}+\sum_{j=1,2}[g_{p,j}(\hat p_{j}^{\dag}\hat a_{cj}+\hat p_{j}\hat a_{cj}^{\dag})]
\\&\hat{H}_{LTC}=\sum_{j=1,2}[\omega_{c,j}\hat a_{cj}^{\dag}\hat a_{cj}]+g_{c,induc}(\hat a_{c1}+\hat a_{c1}^{\dag})(\hat a_{c2}+\hat a_{c2}^{\dag})
\\&\quad\quad\quad\quad+g_{c,cap}(\hat a_{c1}-\hat a_{c1}^{\dag})(\hat a_{c2}-\hat a_{c2}^{\dag})
\end{aligned}
\end{equation}
where $\hat p_{j}=|k\rangle\langle l|_{j}$ ($p_{j}^{\dag}$) is the lowering (raising)
operator for the transition $|k\rangle\leftrightarrow|l\rangle$ with
frequency $\omega_{p,j}$, $a_{cj}$ ($a_{cj}^{\dag}$) denotes the destroy (creation)
operator for the coupler mode with transition frequency $\omega_{c,j}$, and $g_{p,j}=J_{cj} \langle k1|\hat n_j \hat n_{cj}|l0\rangle$
represents the fluxonium-coupler interaction strength.

Here, as mentioned before, we assume degenerate modes for the two fundamental resonator modes within the
LTC, i.e., $\omega_{c,j} = \omega_c$. In addition, the interaction strength $g_{p,j}$ is significantly smaller
than the fluxonium-coupler detuning $\Delta_{p,j} = \omega_{p,j} - \omega_c$ (i.e., the dispersive regime).
Following the procedure given in
Ref.~\cite{Zhao2025} (for easy reference, see also Appendix~\ref{App_A}), an effective system Hamiltonian can be derived by first diagonalizing the LTC
Hamiltonian $\hat{H}{\text{LTC}}$ and then eliminating the direct fluxonium-coupler interactions.
This yields
\begin{equation}
\begin{aligned}\label{eq25}
&\hat{H}_{p}^{({\rm eff})}= \sum_{j=1,2}[\omega_{p,j}\hat p_{j}^{\dag}\hat p_{j}]+g_{p,{\rm eff}}(\hat p_{1}^{\dag}\hat p_{2}+\hat p_{1}\hat p_{2}^{\dag}),
\end{aligned}
\end{equation}
where $g_{p,{\rm eff}}$ is the strength of the coupler-mediated flip-flop interaction between the transition
$|k\rangle \leftrightarrow |l\rangle$ of the two fluxonium qubits, given by
\begin{equation}
\begin{aligned}\label{eq26}
&g_{p,{\rm eff}}=\frac{g_{p,1}g_{p,2}g_{c}}{2}\sum_{j=1,2}\left(\frac{1}{\Delta_{p,j}^{2}}
+\frac{1}{\omega_{c}\Delta_{p,j}}\right).
\end{aligned}
\end{equation}
Note that in deriving the above effective Hamiltonian, the capacitive inter-resonator coupling
terms in Eq.~(\ref{eq24}) are omitted, and we have taken $g_c = g_{c,induc}$. This omission is
justified because the effective capacitance of the resonator is far larger than the coupling
capacitance, thereby heavily suppressing the capacitive coupling strength, as illustrated
in Sec.~\ref{SecIIA2} [see Fig.~\ref{fig3}(b)].

Equation~(\ref{eq26}) demonstrates that when inductive coupling dominates, the two-fluxonium interaction
can be dynamically controlled by tuning the inter-resonator inductive coupling $g_{c,induc}$, which
is achieved by applying a flux bias to the rf-SQUID within the LTC [see Eq.~(\ref{eq18})]. We note that stray
capacitances in the circuit introduce stray capacitive coupling, and in this design, the greatest contribution
may come from the small inherent capacitance of the coupling junction. However, unlike in the DTC~\cite{Zhao2025}, the
capacitive inter-mode coupling here is heavily suppressed due to the large effective capacitance of the mode within
the LTC, as shown in Fig.~\ref{fig3}(b). Consequently, regardless of the coupled qubit
parameters (i.e., independent of the details of the fluxonium transition $|k\rangle \leftrightarrow |l\rangle$), the
coupling nulling condition is determined solely by the LTC itself, as shown in Eq.~(\ref{eq22}). Thus, in addition
to enabling long-distance interaction, the LTC offers a prominent feature, namely an intrinsic nulling condition, that
distinguishes it from existing coupling designs for superconducting qubits, in which different interactions are
generally turned off at different bias conditions~\cite{Zhao2025}.

Before concluding this subsection, we note that while the preceding analytical analysis is carried out
under the dispersive approximation, the implementation of fast gate operations generally necessitates
operation outside this regime, favoring instead the strong non-dispersive regime~\cite{Sung2021,Zhao2025}.
However, operating in the strong non-dispersive regime requires substantial attention to the impact of the induced strong state
hybridization between qubits and couplers. Besides introducing an additional decoherence channel (coupler-induced
decoherence, see, e.g., Ref.~\cite{Zhao2025}), such strong state
hybridization could give rise to several detrimental effects.First, it could affect single-qubit addressing and consequently compromise
single-qubit operations, such as single-qubit gates~\cite{Goerz2017} and qubit readout~\cite{Khezri2015}. Second, it could also impair
the functionality of the coupler design in large-scale systems. In particular, strong state hybridization
arising from spectator (idle) couplers could significantly affect the operation of the target
qubit-coupler-qubit system~\cite{Zajac2021,Zhao2025}. To this end, a tunable coupler with an intrinsic nulling
condition, such as the LTC presented here, could be highly effective at suppressing spectator errors in
larger systems, even when operated in the strong non-dispersive regime~\cite{Zhao2025}.

\subsubsection{Engineering two-fluxonium interactions: A specific example}\label{SecIIB2}

\begin{table}[!htb]
\caption{\label{tab:fluxonium_parameters} Circuit Hamiltonian parameters of the coupled fluxonium system shown
in Fig.~\ref{fig1}(b), considering only the fundamental modes of the two $\lambda/4$ CPW resonators within the LTC. Here, the terminated
linear inductor of the CPW resontaor is $L_{s}=0.5\,\rm nH$ and the detailed CPW parameters are listed in Table~\ref{tab:CPW_parameters}.}
\begin{ruledtabular}
\begin{tabular}{cccc}
(GHz)&
$E_C/2\pi$ &
$E_L/2\pi$ &
$E_J/2\pi$ \\\hline
Fluxonium $Q_{1}$ & 1.00 & 0.80 & 6.55 \\
Fluxonium $Q_{2}$ & 1.00 & 0.75 & 6.60  \\
Resonator $C$ & 0.0403 & 74.7550 & $-$ \\
\hline
\hline
(GHz) & $J_{c1}/2\pi$  & $J_{c2}/2\pi$  & $E_{J,c}/2\pi$ \\\hline
Coupling strengths & 0.125 & 0.125  & 99.336
\end{tabular}
\end{ruledtabular}
\end{table}

Following the above analytical analysis, we now provide a more explicit analysis based on a specific implementation
of fluxonium qubits coupled via a one-centimeter LTC (F-LTC-F), as shown in Fig.~\ref{fig1}(b). The system parameters used
in this analysis are summarized in Table~\ref{tab:fluxonium_parameters}. Before proceeding with the explicit analysis, it
is worth clarifying the rationale behind these parameter
choices. Our goal is twofold. First, we aim to construct a long-range coupler capable of connecting fluxonium
qubits separated by, for illustration purposes, one centimeter. Second, we seek to achieve high-contrast
control over the fluxonium interaction associated with the plasmon transition $|1\rangle \leftrightarrow |2\rangle$.
This high-contrast control, in turn, would enable the suppression of quantum crosstalk~\cite{Zhao2025,Zhao2026,Zwanenburg2026} and
facilitate sub-100-ns CZ gates by selectively activating such a transition, as demonstrated in
Refs.~\cite{Ding2023,Nesterov2018,Ficheux2021,Zhao2025}.

\begin{figure}[tbp]
\begin{center}
\includegraphics[keepaspectratio=true,width=\columnwidth]{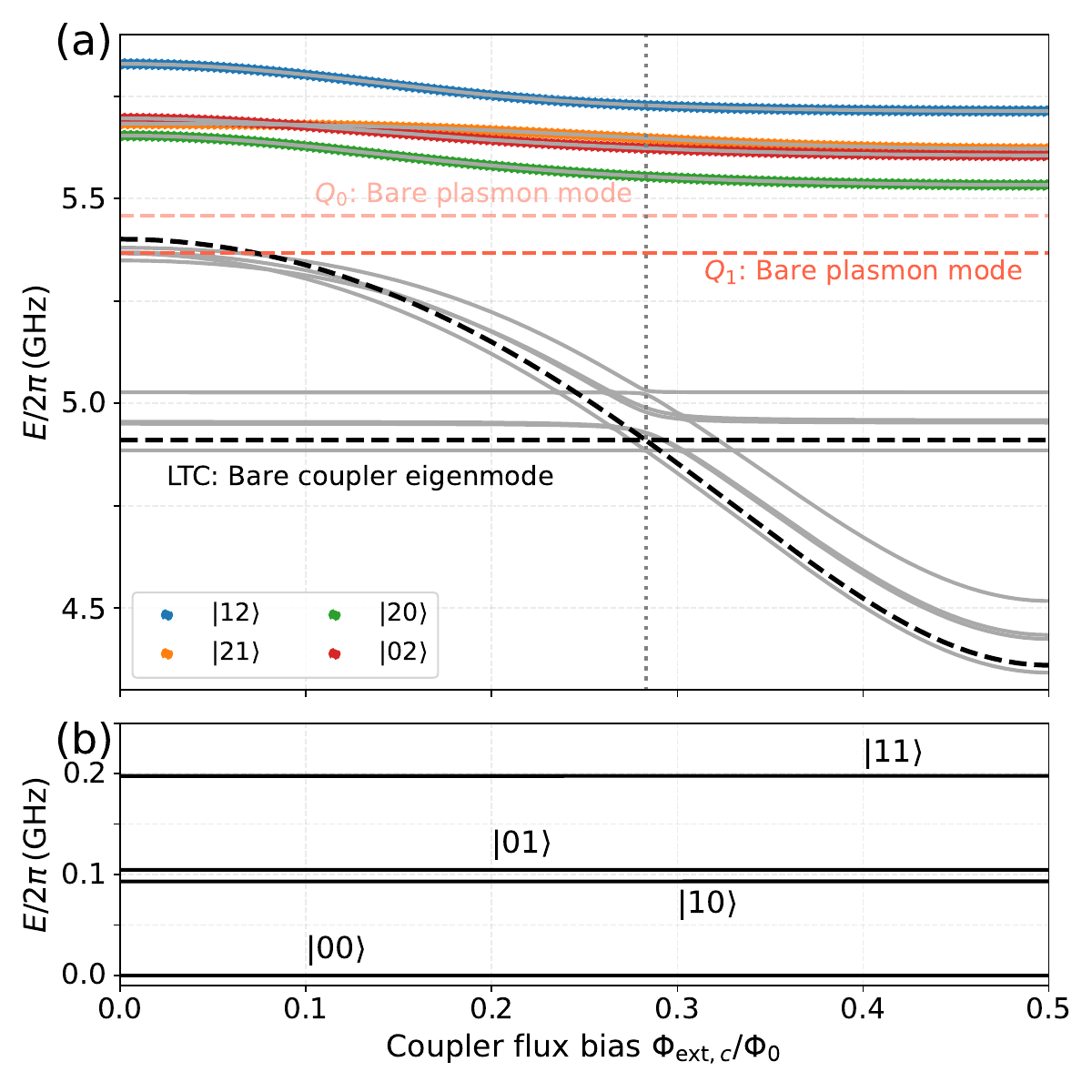}
\end{center}
\caption{Energy levels of the F-LTC-F coupled system versus the coupler flux bias.
(a) Noncomputational energy levels associated with the plasmon transition $|1\rangle \leftrightarrow |2\rangle$ of the full
coupled system, along with the LTC's bare eigenmode frequencies (excluding coupling to the fluxonium qubits), plotted as functions
of the coupler flux bias. The horizontal dashed lines indicate the bare plasmon transition
frequencies ($|1\rangle \leftrightarrow |2\rangle$) of the fluxonium qubits, and the vertical dashed line marks the
flux bias at which the LTC's inter-resonator inductive coupling is turned off (excluding coupling to the fluxonium qubits).
(b) Computational energy levels $\{|00\rangle,\,|01\rangle,\,|10\rangle,\,|11\rangle\}$ versus the coupler flux bias.}
\label{fig4}
\end{figure}

To realize this design, we adopt a specific LTC configuration based on the CPW parameters summzrized in
Table~\ref{tab:CPW_parameters}. As illustrated in Sec.~\ref{SecIIA} and shown in Fig.~\ref{fig2}(d), the one-centimeter LTC
consists of two identical $\lambda/4$ CPW resonators terminated with the same linear inductor $L_{S} = 0.5\,\rm nH$. For this
configuration, the degenerate mode frequency $\omega_{c}$ of the LTC is approximately $4.9\,\rm GHz$ [see the dashed
black line in Fig.~\ref{fig4}(a)]. Furthermore, according to Eq.~(\ref{eq26}), achieving large fluxonium interactions generally requires a
small fluxonium-coupler detuning and strong fluxonium-coupler couplings. To satisfy this requirement, we choose
to place the frequency of the $|1\rangle \leftrightarrow |2\rangle$ transition above the coupler mode
frequency [see the horizontal dashed lines in Fig.~\ref{fig4}(a)]. Note here that, on one hand, a smaller detuning
is highly desirable for achieving strong coupling and fast gate speed. On the other hand, one should avoid introducing
additional error channels or affecting other qubit operations, such as single-qubit gates and readout (We will return to
this issue in a later section).

Given the parameters in Table~\ref{tab:fluxonium_parameters}, Fig.~\ref{fig4} shows the energy spectrum of the F-LTC-F system as a
function of the coupler flux bias applied to the rf-SQUID. Note that, while the inter-resonator capacitive coupling
with $C_{J} = 2.5\,\rm fF$ is omitted in the above analytical analysis, this capacitive coupling is included in the
following numerical analysis. For easy reference, the LTC's bare eigenmode frequencies and the bare
frequencies of the plasmon transition $|1\rangle \leftrightarrow |2\rangle$ are also shown. As mentioned in Sec.~\ref{SecIIB1}, the
transition dipole of the qubit transition is typically much smaller. Consequently, the computational energy levels exhibit almost
negligible dependence on the coupler flux bias~\cite{Zhao2025}, as seen in Fig.~\ref{fig4}(b). In contrast, the transmon-like large dipole of the
plasmon transition $|1\rangle \leftrightarrow |2\rangle$ leads to its strong coupling with the LTC. This is evidenced
by the strong level shift between the bare energy levels and the full system energy levels. As a result, this strong coupling
can be engineered to give rise to significant coupler-mediated fluxonium interactions for the plasmon
transition $|1\rangle \leftrightarrow |2\rangle$. Additionally, the zero-coupling bias, identified as the bias at which the two
bare eigenmodes of the LTC become degenerate, is approximately $0.283$. This value agrees well with the expression given
in Eq.~(\ref{eq22}), even when accounting for the inter-resonator capacitive coupling.

To more explicitly characterize the LTC-mediated high-contrast interaction, we follow the definition introduced in
Ref.~\cite{Zhao2025} and consider two complementary metrics: state hybridization and state-dependent frequency shifts.
State hybridization is quantified by the overlap between the full system eigenstates $|kl\rangle$ and the corresponding (coupled) bare
states $|\overline{lk}\rangle$. For example, $|\langle 12|\overline{21}\rangle|^{2}$ (or $|\langle 21|\overline{12}\rangle|^{2}$) quantifies
the LTC-mediated coupling for the transition $|1\rangle \leftrightarrow |2\rangle$, i.e., the
interaction $|12\rangle \leftrightarrow |21\rangle$. The state-dependent frequency shifts are defined as follows, for
the $|1\rangle \leftrightarrow |2\rangle$ transition of $Q_1$ and $Q_2$, respectively: $\delta\omega_{12,i}=|(E_{21}-E_{11})-(E_{20}-E_{10})|$ and $\delta\omega_{i,12}=|(E_{12}-E_{11})-(E_{02}-E_{01})|$, where $E_{kl}$ denotes the energy of the eigenstate $|kl\rangle$.

\begin{figure}[tbp]
\begin{center}
\includegraphics[keepaspectratio=true,width=\columnwidth]{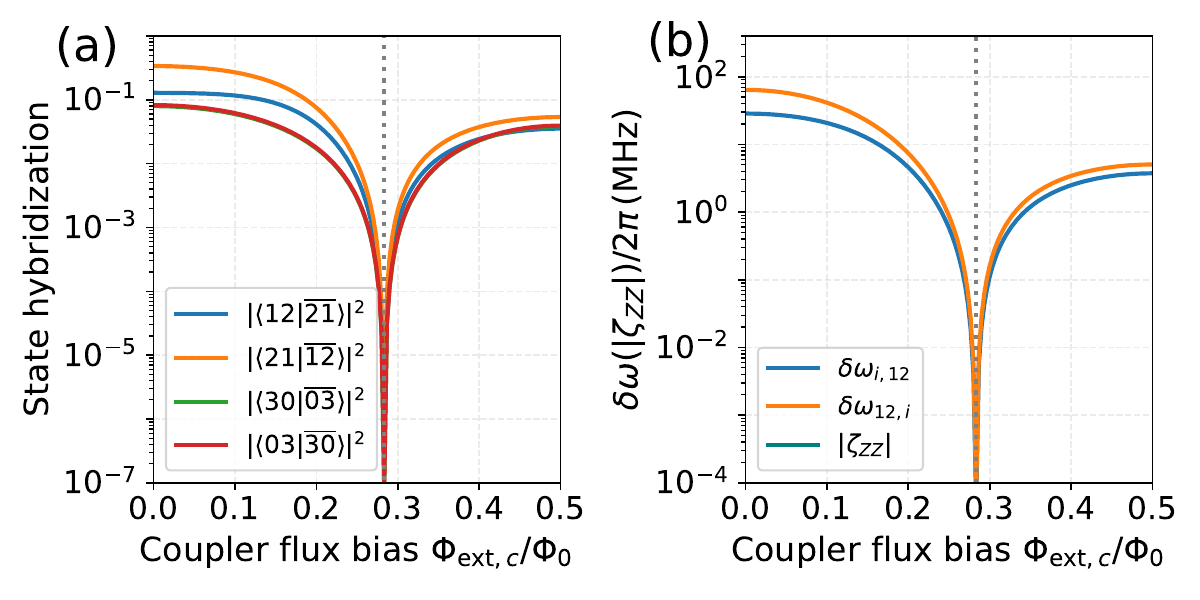}
\end{center}
\caption{LTC-mediated fluxonium interactions quantified by state hybridization and state-dependent
frequency shifts. (a) State hybridization corresponding to the LTC-mediated interactions for the
plasmon transitions $|1\rangle \leftrightarrow |2\rangle$ and $|0\rangle \leftrightarrow |3\rangle$ as
functions of the coupler flux bias. The vertical dashed line marks the flux bias at which the state
hybridization is minimized. (b) State-dependent frequency shifts for the $|1\rangle \leftrightarrow |2\rangle$
plasmon transition of the two fluxonium qubits versus the coupler flux bias. The vertical dashed line marks
the flux bias at which the shift is minimized. The ZZ coupling $\zeta_{ZZ}$ is also plotted and is found to
remain below $0.1\,\rm kHz$ over the entire bias range.}
\label{fig5}
\end{figure}

Accordingly, Figures~\ref{fig5}(a) and~\ref{fig5}(b) show the state hybridization and the state-dependent
frequency shifts associated with the $|1\rangle \leftrightarrow |2\rangle$ transition as functions of the
coupler flux bias. In Fig.~\ref{fig5}(a), the state hybridization for the LTC-mediated coupling for
the $|0\rangle \leftrightarrow |3\rangle$ transition is also shown. As expected, the state hybridizations for
both plasmon transitions, and consequently the corresponding LTC-mediated couplings, are simultaneously minimized
at a bias of approximately $0.283$ and maximized at a bias of $0$. These results agree well with the analytical
analysis presented in the previous section. Furthermore, this observation also applies to the state-dependent frequency
shifts shown in Fig.~\ref{fig5}(b). Here, the ZZ coupling, defined as $\zeta_{ZZ}=(E_{11}-E_{01})-(E_{10}-E_{00})$~\cite{DiCarlo2009}, is
also plotted and is found to remain below $0.1,\rm kHz$ over the entire bias range. This confirms the decoupling of
qubit states, which is attributed to the small transition dipole and the low qubit transition frequency.

\begin{figure}[tbp]
\begin{center}
\includegraphics[keepaspectratio=true,width=\columnwidth]{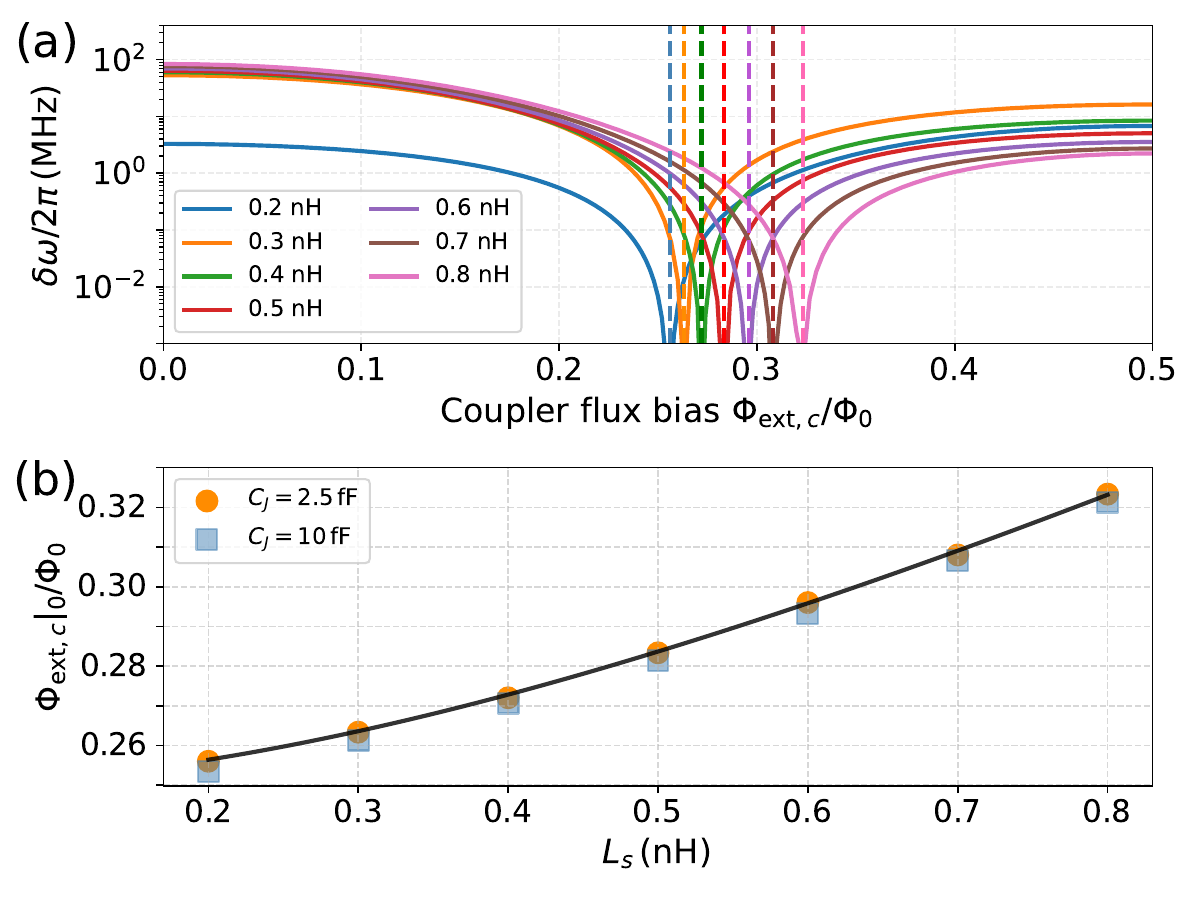}
\end{center}
\caption{The dependence of LTC-mediated fluxonium interactions on the terminating linear inductor $L_{S}$.
(a) LTC-mediated fluxonium interactions, quantified by the state-dependent frequency shift $\delta\omega_{12,i}$, as
a function of the flux bias for varying $L_{S}$. Vertical dashed lines indicate the zero-coupling points, where
the shift is minimized. (b) Numerical and analytical results for the zero-coupling points as a function of the
terminating linear inductor $L_{S}$. For the numerical results, we consider both the case with a junction
capacitance of $2.5\,\rm fF$, as in (a), and the case with $10\,\rm fF$.}
\label{fig6}
\end{figure}

Along with the analytical analysis, the above numerical analysis demonstrates that the LTC enables a coupler
design with an intrinsic nulling condition. To further validate this point and to study the dependence of the
LTC-mediated coupling on the terminating linear inductor, Figure~\ref{fig6}(a) shows the state-dependent
frequency shift $\delta\omega_{12,i}$ as a function of the flux bias for varying $L_{S}$, with the
zero-coupling points indicated by the vertical dashed lines (where the shift is minimized).
Generally, a larger $L_{S}$ is found to enable stronger coupling between fluxonium qubits. This behavior is
expected, as a larger $L_{s}$ generally gives rise to a larger $\eta$ [see Fig.~\ref{fig2}(e)] and thus enhances the
inter-resonator inductive couplings [see Eqs.~(\ref{eq18}) and~(\ref{eq26})].

Given the results shown in Fig.~\ref{fig6}(a), Figure~\ref{fig6}(b) plots the dependence of the zero-coupling
points on the terminating linear inductor $L_{S}$. Moreover, the analytical results for the zero-coupling
points, derived from Eq.~(\ref{eq22}), are also shown. As expected, the analytical results show excellent
agreement with the numerical results, which account for the junction capacitance of $2.5\,\rm fF$. To further
examine the limit of this agreement, we also present numerical results with $C_{J}$ increased to $10\,\rm fF$, which
still align well with the analytical results. These analyses confirm that inter-resonator inductive coupling
is the dominant coupling mechanism in the considered circuit, and that the zero-coupling point is determined
solely by the LTC itself. Overall, this behavior is attributed to the aforementioned fact that the effective
capacitance of the resonator mode within the LTC is far larger than the junction capacitance assumed here,
thereby heavily suppressing inter-resonator capacitive couplings.

\section{CZ gate realization}\label{SecIII}

Based on the F-LTC-F system studied in the previous section and the system parameters listed in
Table~\ref{tab:fluxonium_parameters}, here we first present the implementation of
the CZ gate on the two fluxonium qubits and analyze its dominant
intrinsic error sources. We then extend the discussion to scenarios involving a longer coupler
and examine the CZ gate performance under such conditions. Finally, based on the aforementioned
analysis, we address the following question: how much coupling strength $J_{cj}$ is needed for a sub-100-ns CZ
gate within the F-LTC-F architecture?

\subsection{Microwave-activated CZ gates: implementation and error analysis}\label{SecIIIA}

As shown in Fig.~\ref{fig5}(b), the LTC-mediated interactions (for the transition $|1\rangle \leftrightarrow |2\rangle$)
cause the plasmon transition frequency of a given fluxonium qubit to depend on the state of its coupled neighbors. Such
state-dependent frequency shifts enable the selective excitation of transitions such as $|10 (11)\rangle \leftrightarrow |20 (21)\rangle$
and $|01(11)\rangle \leftrightarrow |02(12)\rangle$. Consequently, CZ gates can be realized by first tuning the LTC from its
idle point (zero-coupling point; here we set $\varphi_{\text{ext},c}/2\pi = 0.283$) to the interaction point (where the plasmon interaction is turned on; here we set $\varphi_{\text{ext},c}/2\pi = 0$, giving rise to the frequencies $5.48556\,(5.54947)\,\rm GHz$
and $5.60279\,(5.63136)\,\rm GHz$ for the former four transitions), then applying
a microwave drive to the fluxonium qubits and waiting for one period of the selectively driven Rabi
oscillation (during which the system accumulates a conditional phase factor of $\pi$), and finally biasing
the LTC back to the idle point. Note here that due to the near-complete decoupling of the computational subspace
from the coupler (see, e.g., Fig.~\ref{fig4}(b) and Ref.~\cite{Zhao2025}), non-adiabatic transitions for qubit
states during the flux bias ramping can be neglected. Thus, the following discussion excludes the coupler
bias ramping.

Following the approaches in Refs.~\cite{Nesterov2018,Ding2023,Zhao2025}, we here employ simultaneous microwave
drives with identical amplitude and frequency applied to both $Q_1$ and $Q_2$ to activate the target gate transition
and to reduce unwanted phase accumulations from off-resonant transitions. The corresponding driven Hamiltonian takes the form
\begin{equation}
\begin{aligned}\label{eq27}
H_{d}=\sum_{j=1,2}A(t)\cos(\omega_{d}t+\phi_{j})\hat n_{j},
\end{aligned}
\end{equation}
where $A$, $\omega_d$, and $\phi_j$ denote the amplitude, frequency, and phase of the drive, respectively. The relative
phase between the two drives is determined by maximizing constructive interference at the target gate
transition~\cite{Ding2023}.

Here, to mitigate the nearest spurious transition relative to the target gate transition, a cosine-shaped
microwave drive combined with the DRAG scheme~\cite{Motzoi2009} is utilized (see Appendix~\ref{App_B}). As shown in
Fig.~\ref{fig5}(b), for $CZ_{|11\rangle \leftrightarrow |21\rangle}$, the nearby spurious transition
is $|10\rangle \leftrightarrow |20\rangle$, which lies $|\delta\omega_{12,i}|/2\pi=63.92,\rm MHz$ above the gate transition
$|11\rangle \leftrightarrow |21\rangle$; for $CZ_{|11\rangle \leftrightarrow |12\rangle}$, the nearby
spurious transition is $|01\rangle \leftrightarrow |02\rangle$, which lies $|\delta\omega_{i,12}|/2\pi=28.57\,\rm MHz$ below the
gate transition $|11\rangle \leftrightarrow |12\rangle$. In addition, the gate drive parameters, including
the drive amplitude and frequency, are determined by minimizing both the conditional phase error (deviation of
the accumulated phase from $\pi$) and the population leakage outside the computational subspace (spanned
by ${|00\rangle, |01\rangle, |10\rangle, |11\rangle}$).

\begin{figure}[tbp]
\begin{center}
\includegraphics[keepaspectratio=true,width=\columnwidth]{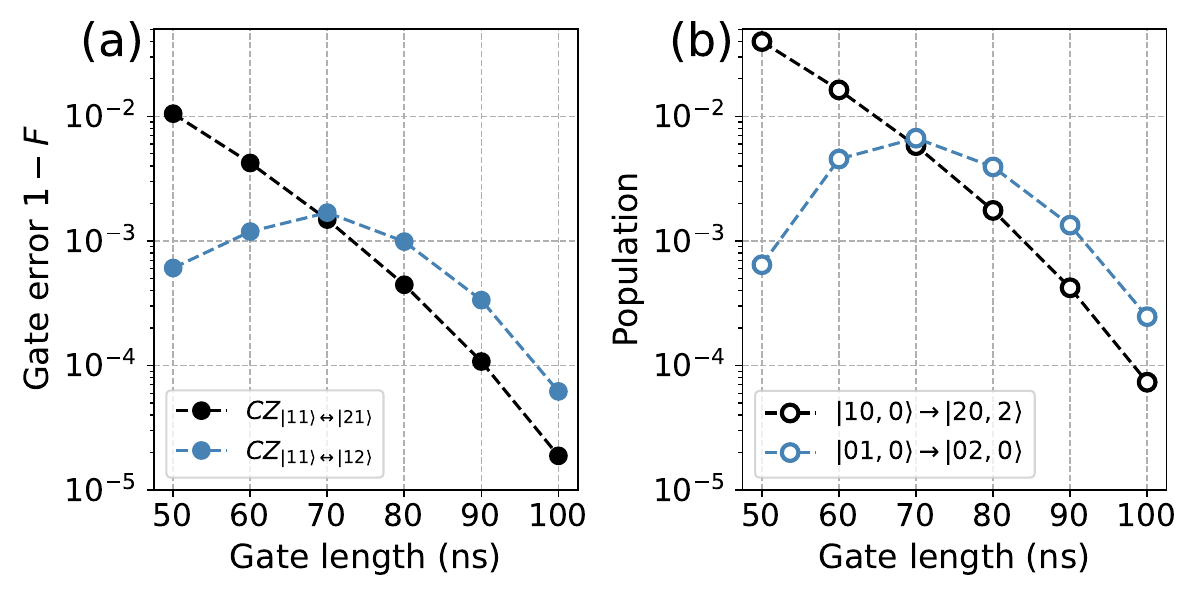}
\end{center}
\caption{Intrinsic gate error of microwave-activated CZ gates and the dominant intrinsic error sources. Here, during the
gate operations, the LTC is tuned from its idle point [$\varphi_{\text{ext},c}/(2\pi) = 0.283$] to the interaction
point [$\varphi_{\text{ext},c}/(2\pi) = 0$], held for the duration of the microwave drive, and then returned.
A cosine-shaped microwave drive combined with the DRAG scheme is used to mitigate the most nearby spurious transition.
(a) Intrinsic gate error as a function of gate length for CZ gate implementation by activating the $|11\rangle \leftrightarrow |21\rangle$ transition ($CZ_{|11\rangle \leftrightarrow |21\rangle}$) and the $|11\rangle \leftrightarrow |12\rangle$ transition ($CZ_{|11\rangle \leftrightarrow |12\rangle}$).
(b) The dominant intrinsic error sources for the CZ gate implementations, specifically $|10,0\rangle \leftrightarrow |20,2\rangle$ for $CZ_{|11\rangle \leftrightarrow |21\rangle}$ and $|01,0\rangle \leftrightarrow |02,0\rangle$ for $CZ_{|11\rangle \leftrightarrow |12\rangle}$. }
\label{fig7}
\end{figure}

Following the above settings, the intrinsic gate error~\cite{Pedersen2007} (see Appendix~\ref{App_B}) as a function of gate length is shown in
Fig.~\ref{fig7}(a), demonstrating that sub-100-ns CZ gates can be achieved with an intrinsic gate
error below $10^{-4}$. As expected, increasing the gate length generally suppresses the gate error.
This trend holds particularly well for the $CZ_{|11\rangle \leftrightarrow |21\rangle}$ case, whereas
for $CZ_{|11\rangle \leftrightarrow |12\rangle}$, clear oscillations in the gate error are observed.

To illustrate the mechanism behind this difference, we examine the system dynamics during both gates.
Figure~\ref{fig7}(b) shows the dominant intrinsic error sources for $CZ_{|11\rangle \leftrightarrow |21\rangle}$
and $CZ_{|11\rangle \leftrightarrow |12\rangle}$. One can see that for $CZ_{|11\rangle \leftrightarrow |21\rangle}$, the
DRAG scheme effectively mitigates the activation of the nearby transition, leaving the high-order two-photon
process $|10,0\rangle \leftrightarrow |20,2\rangle$ as the dominant source of intrinsic gate error.
In contrast, for $CZ_{|11\rangle \leftrightarrow |12\rangle}$, even with the DRAG scheme, the nearby
spurious transition $|01,0\rangle \leftrightarrow |02,0\rangle$ remains the dominant error source.
This is expected, as the state-dependent frequency shift for the latter is
only $|\delta\omega_{i,12}|/2\pi = 28.57\,\rm MHz$, while for the former it
is $|\delta\omega_{12,i}|/2\pi = 63.92\,\rm MHz$.

The above observation explains why the gate error depends differently on gate length. For $CZ_{|11\rangle \leftrightarrow |21\rangle}$, which
shows a monotonic error decrease without oscillations, the higher-order (two-photon) $|10,0\rangle \rightarrow |20,2\rangle$ transition
induces only slow, weak population leakage oscillations. For $CZ_{|11\rangle \leftrightarrow |12\rangle}$, by contrast, direct off-resonant
driving of the $|01,0\rangle \leftrightarrow |02,0\rangle$ transition generates fast, strong
oscillating population leakage, as shown in Fig.~\ref{fig7}(b).

\begin{figure}[tbp]
\begin{center}
\includegraphics[keepaspectratio=true,width=\columnwidth]{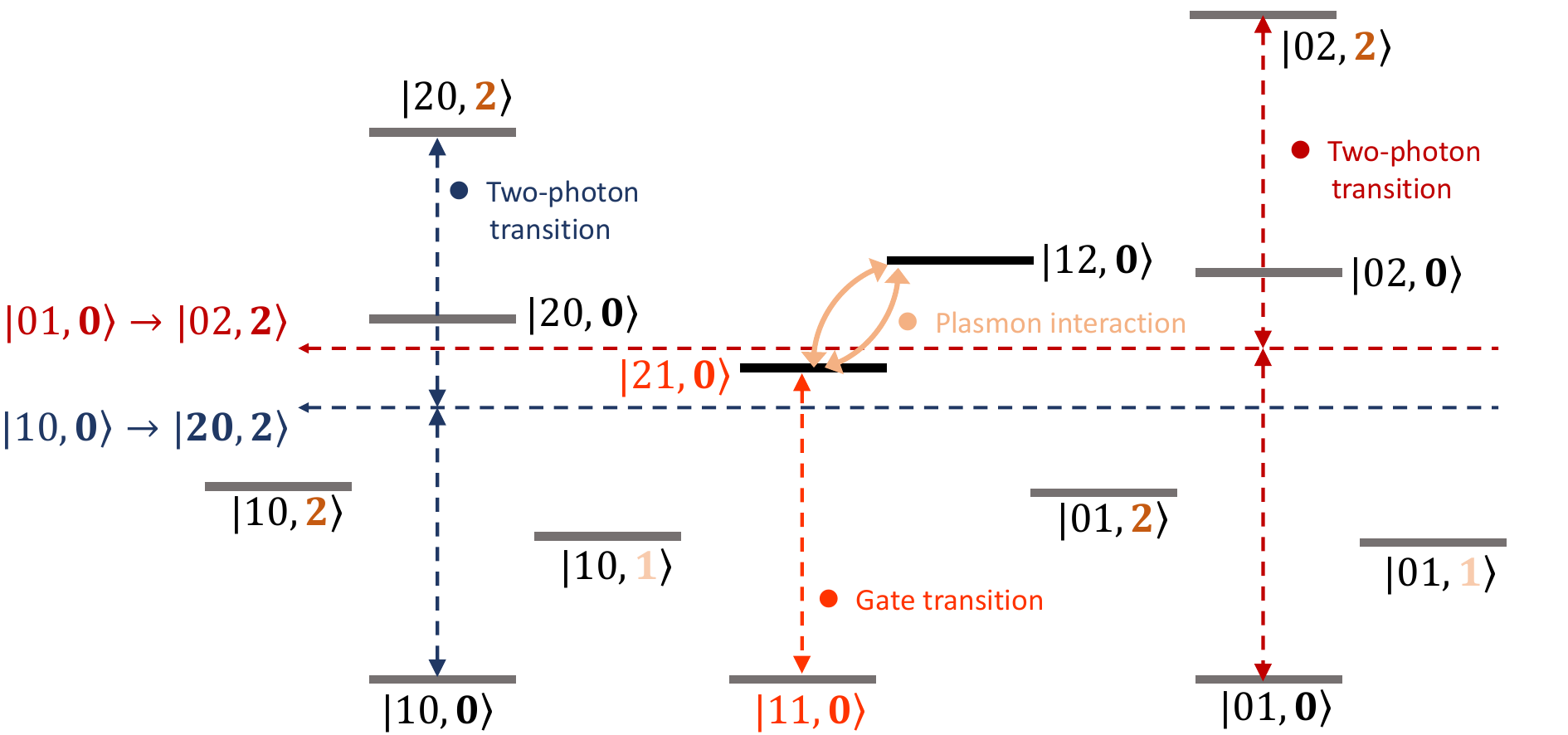}
\end{center}
\caption{Schematic illustrating the mechanism of two-photon transitions that limit microwave-activated CZ gates in
the F-LTC-F system. Under the current frequency allocation, the frequency of the $|1\rangle \leftrightarrow |2\rangle$ plasmon
transition is positioned above the coupler mode frequency. This configuration may give rise to two potential frequency collisions
associated with the two-photon transitions $|10,0\rangle \leftrightarrow |20,2\rangle$ and $|01,0\rangle \leftrightarrow |02,2\rangle$.}
\label{fig8}
\end{figure}

To further reduce the intrinsic gate error in the F-LTC-F system, particularly when the DRAG
scheme cannot efficiently suppress a nearby spurious transition (as in the case of $CZ_{|11\rangle \leftrightarrow |12\rangle}$),
one can adopt the synchronized gate scheme demonstrated in Ref.~\cite{Ficheux2021}. For gates limited by
higher-order transitions, such as $CZ_{|11\rangle \leftrightarrow |21\rangle}$, a more careful allocation
of qubit frequencies is required, guided by a systematic analysis of frequency collision conditions for
higher-order transitions. Here, we provide a systematic analysis of the latter case, while leaving the
former case to be discussed in the following subsection, where we consider an even smaller frequency
shift of $|\delta\omega|/2\pi \sim 10\,\rm MHz$.

As shown in Fig.~\ref{fig4}(a), in the current F-LTC-F system, the frequency of the $|1\rangle \leftrightarrow |2\rangle$ plasmon
transition is positioned above the coupler mode frequency. Consequently, this configuration can give rise to two frequency collisions
associated with the two-photon transitions $|10,0\rangle \leftrightarrow |20,2\rangle$ and $|01,0\rangle \leftrightarrow |02,2\rangle$,
as illustrated in Fig.~\ref{fig8}. While such two-photon transitions are allowed in any coupled system, they are generally
negligible when the coupler frequency is far detuned from the fluxonium (qubit) transition (note that this case indeed
corresponds to the two-photon transition associated with the lower eigenmodes of the LTC, which is omitted here; we
concern ourselves only with the case involving the higher eigenmodes of the LTC). However, when the coupler-qubit detuning
becomes smaller, these transitions should be carefully taken into consideration, as noted in Ref.~\cite{Zhao2023} and also
relevant to our present system. Generally, the direct approach to mitigate their effect on gates is to put the system away from these
frequency collision points.

\begin{figure}[tbp]
\begin{center}
\includegraphics[keepaspectratio=true,width=\columnwidth]{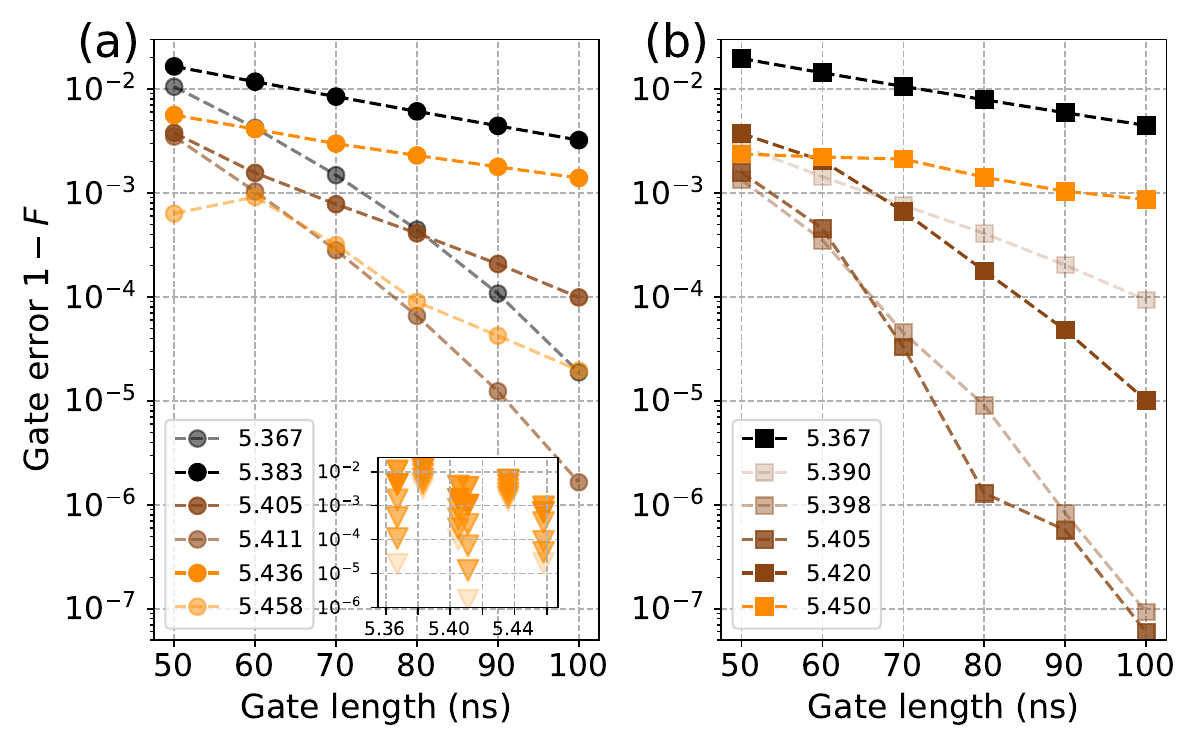}
\end{center}
\caption{Intrinsic gate error of $CZ_{|11\rangle \leftrightarrow |21\rangle}$ in the F-LTC-F system with varying plasmon
transition frequencies. Other system parameters and control settings are the same as in Fig.~\ref{fig7}.
(a) Intrinsic CZ gate error as a function of gate length for varying plasmon transition frequencies of $Q_1$, while
the plasmon transition frequency of $Q_2$ is fixed at approximately $90\,\rm MHz$ above that of $Q_1$.
(b) Intrinsic CZ gate error as a function of gate length for varying plasmon transition frequencies of $Q_1$, while the
plasmon transition frequency of $Q_2$ is fixed at approximately $120\,\rm MHz$ above that of $Q_1$.}
\label{fig9}
\end{figure}

To more explicitly study the effects of these collisions on gates, Figure~\ref{fig9}(a) shows the intrinsic error of
$CZ_{|11\rangle \leftrightarrow |21\rangle}$ as a function of gate length for varying plasmon transition frequencies
of $Q_1$, while the plasmon transition frequency of $Q_2$ is fixed at approximately $90\,\rm MHz$ above that of $Q_1$ (full fluxonium qubit
parameters can be found in Appendix~\ref{App_C}; other system parameters and control settings are the same as in Fig.~\ref{fig7} and are also listed in Table~\ref{tab:fluxonium_parameters}). Similar to Fig.~\ref{fig7}(a), Figure~\ref{fig9}(a) shows a monotonic decrease in
error with increasing gate length, regardless of the plasmon transition frequency. More notably, when the plasmon transition frequencies
approach two specific frequency points, the decreasing trend becomes flatter and smoother, revealing two error
hotspots [see the inset in Fig.~\ref{fig9}(a)]. This suggests that the F-LTC-F system operates near the aforementioned two frequency
collision points.

By contrast, when the system is operated below the lower collision frequency, above the higher collision
frequency, or in the middle of the two collision frequencies, the decreasing trend of the gate error becomes steeper. The same
observation can also be made when the plasmon transition frequency of $Q_2$ is fixed at approximately $120\,\rm MHz$ above
that of $Q_1$ (full fluxonium parameters can be found in Appendix~\ref{App_C}), as shown in Fig.~\ref{fig9}(b).

\subsection{CZ gate realization with extended coupler lengths}\label{SecIIIB}

Here, we further consider implementing CZ gates in the F-LTC-F system with an even longer LTC and
evaluate the gate performance, thereby providing an intuitive view of the main challenges for further
extending the LTC distance. Following the discussion in Sec.~\ref{SecIIA} and considering the CPW parameters
listed in Table~\ref{tab:CPW_parameters}, increasing the length of each $\lambda/4$ CPW resonator
from $0.5\,\rm cm$ to $0.75\,\rm cm$ leads to the following changes: (i) the fundamental mode frequency
decreases from approximately $4.9\,\rm GHz$ to $3.6\,\rm GHz$, and the first harmonic mode frequency
decreases from approximately $15\,\rm GHz$ to $11\,\rm GHz$; (ii) the effective capacitance of the fundamental
mode increases from approximately $480\,\rm fF$ to $680\,\rm fF$ (lowering the resonator impedance and increasing the
zero-point fluctuation of charge number).

Regarding the former point, further increasing the length, and thus further decreasing the
harmonic mode frequency, could degrade the validity of the present model. In such cases, the impact
of these higher harmonic modes on the F-LTC-F system should not be neglected. Moreover, to ensure
fast two-qubit gates, the fluxonium circuit parameters would need to be adjusted accordingly so that the
plasmon frequency remains not far detuned from the LTC modes. This becomes challenging when the mode frequency
is further decreased.

For the latter point, the larger capacitance of the resonator will require an even larger
coupling capacitance to maintain the fluxonium-LTC coupling strength, such as $J_{cj}/2\pi = 125\,\rm MHz$ as in
Table~\ref{tab:fluxonium_parameters}. This is a non-trivial task for fluxonium qubits with a total charge energy of
only $\sim1\,\rm GHz$, as a significant portion of the fluxonium's charging energy would then come from such
a larger coupling capacitor. Moreover, a practical quantum device would require each fluxonium qubit to have
multiple coupled neighbors, not even accounting for the contributions from readout and control lines. We will
return to this coupling loading issue in the next section.

Moreover, if keeping $I_c = 200\,\rm nA$ and $L_S = 0.5\,\rm nH$, the maximum inductive inter-resonator coupling
decreases from approximately $250\,\rm MHz$ to $150\,\rm MHz$. As indicated by Eq.~(\ref{eq26}), this reduction
could weaken the LTC-mediated fluxonium interaction and consequently affect the gate speed.

\begin{table}[!htb]
\caption{\label{tab:LTC_F_parameters} The circuit parameters of the F-LTC-F system with an extended
coupler length ($1.5\,\rm cm$) include the updated fluxonium circuit parameters, the effective parameters of the
fundamental mode of the $\lambda/4$ CPW resonator, and the associated CPW parameters, including the
total length $d$, the capacitance $C_l$ and inductance $L_l$ per unit length, and the characteristic
impedance $Z_0$ of the CPW structure. Here, the terminating linear inductor of each resonator
is $L_s = 0.7\,\rm nH$. Other coupling parameters are the same as those in Table~\ref{tab:fluxonium_parameters}.}
\begin{ruledtabular}
\begin{tabular}{cccc}
(GHz)& $E_C/2\pi$ & $E_L/2\pi$ & $E_J/2\pi$ \\\hline
Fluxonium $Q_{1}$ & 1.000 & 0.550 & 4.450 \\
Fluxonium $Q_{2}$ & 1.000 & 0.650 & 4.692  \\
Resonator $C$ & 0.0387 & 38.8814 & $-$ \\
\hline\hline
$ d$ (cm)& $L_l$ (nH/cm)& $C_l$ (fF/cm) & $Z_{0}$ ($\Omega$) \\\hline
0.75 & 5.97 & 1160.67  & 71.71\\
\end{tabular}
\end{ruledtabular}
\end{table}

\begin{figure}[tbp]
\begin{center}
\includegraphics[keepaspectratio=true,width=\columnwidth]{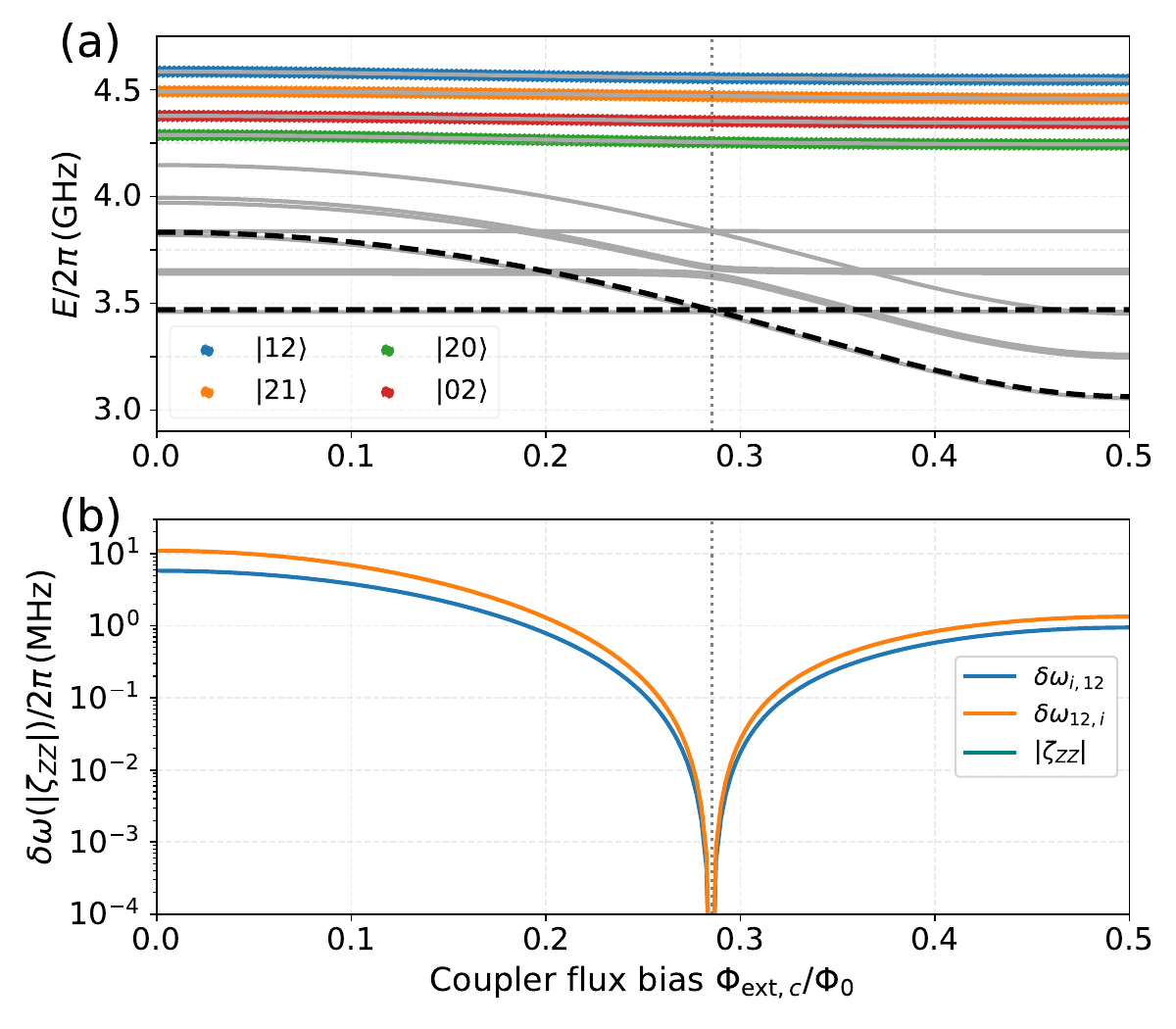}
\end{center}
\caption{System spectrum and LTC-mediated plasmon interactions of the F-LTC-F system with an extended coupler length ($1.5\,\rm cm$).
(a) Non-computational energy levels (for the $|1\rangle \leftrightarrow |2\rangle$ plasmon transition) and LTC bare
eigenmode frequencies versus coupler flux bias. Horizontal dashed lines indicate the bare fluxonium plasmon transition
frequencies. The vertical dashed line marks the bias at which the LTC's inter-resonator inductive coupling is turned
off (coupling to fluxonium excluded).
(b) State-dependent frequency shifts for the $|1\rangle \leftrightarrow |2\rangle$ plasmon transition versus coupler flux
bias. The vertical dashed line marks the shift minimum. The ZZ coupling $\zeta_{ZZ}$ remains below $0.1\,\rm kHz$ across
the entire bias range.}
\label{fig10}
\end{figure}

To ensure potentially practical physical implementations, we here mainly focus on addressing the coupling loading
issue and consider the refined CPW parameters summarized in Table~\ref{tab:LTC_F_parameters}. These parameters
give rise to an effective capacitance of the fundamental resonator mode of approximately $500\,\rm fF$, which remains
at almost the same level as before. In addition, we also consider a larger terminating inductor of $L_S = 0.7\,\rm nH$.
Accordingly, Figure~\ref{fig10}(a) shows the eigenmodes of the LTC with a total length of $1.5\,\rm cm$ and the energy
levels of the F-LTC-F system as a function of the coupler flux bias, using the updated fluxonium parameters also
summarized in Table~\ref{tab:LTC_F_parameters}. Other coupling parameters are the same as those in
Table~\ref{tab:fluxonium_parameters}. Furthermore, as shown in Fig.~\ref{fig10}(b), the LTC-mediated plasmon
interaction for the $|1\rangle \leftrightarrow |2\rangle$ transition, quantified by the state-dependent frequency
shift, is plotted as a function of the coupler flux bias. In agreement with the analytical results given in
Eq.~(\ref{eq22}), the interaction is turned off at approximately $0.285$.

\begin{figure}[tbp]
\begin{center}
\includegraphics[keepaspectratio=true,width=\columnwidth]{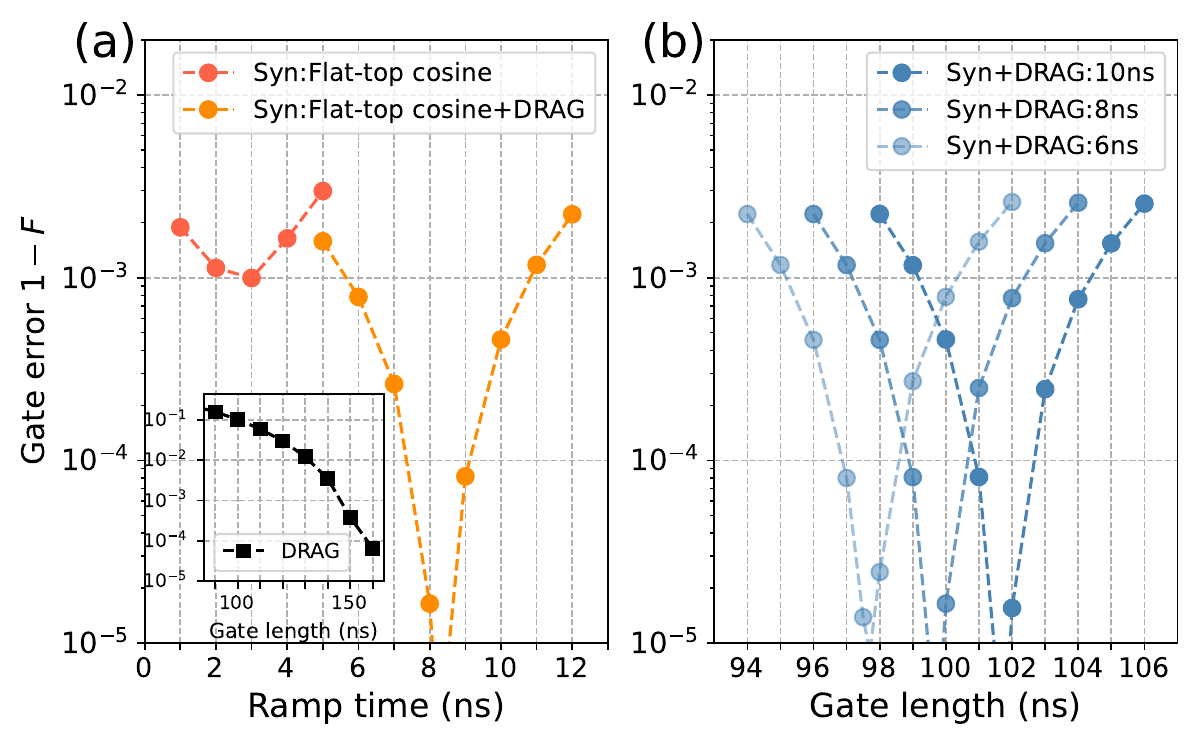}
\end{center}
\caption{Intrinsic gate error of $CZ_{|11\rangle \leftrightarrow |21\rangle}$ in the F-LTC-F system with an extended coupler length ($1.5\,\rm cm$).
Except for the results shown in the inset, where a cosine-shaped pulse is used, a flat-top pulse with cosine ramps is used elsewhere.
(a) Gate error as a function of ramp time for both the synchronized pulse design and the design combining the synchronized scheme with
the DRAG scheme. For the synchronized pulse, the length of the flat-top part is fixed at $2\pi / \delta\omega_{12,i} \approx 90.6\,\rm ns$; for the combined pulse design, the total gate length is fixed at $100\,\rm ns$. The inset shows results for gate implementation using only the cosine-pulse-based DRAG scheme. (b) Gate error for the combined pulse design as a function of gate length for varying ramp times.}
\label{fig11}
\end{figure}

Here, following the gate setting adopted in the preceding subsection, we turn to consider the implementation of
CZ gates in this extended system. At $\varphi_{\text{ext},c}/2\pi = 0$, the transition frequencies are $4.07576$ ($4.08679$) $\rm GHz$
and $4.16270$ ($4.16851$) $\rm GHz$ for $|10 (11)\rangle \leftrightarrow |20 (21)\rangle$ and
$|01 (11)\rangle \leftrightarrow |02 (12)\rangle$, respectively. Since the maximum frequency shift is only about $11.04\,\rm MHz$, the
DRAG scheme is inefficient for suppressing leakage from spurious transitions (as also noted in the preceding subsection), particularly
for short gate lengths. This is further illustrated in the inset of Fig.~\ref{fig11}(a), where a cosine-shaped drive combined with
the DRAG scheme is employed, as in Fig.~\ref{fig7}. One can observe that for $CZ_{|11\rangle \leftrightarrow |21\rangle}$, achieving
an intrinsic gate error below $10^{-4}$ requires a gate length longer than $150\,\rm ns$.

To ensure the implementation of high-fidelity sub-100-ns CZ gates, we alternatively consider the synchronized
gate scheme~\cite{Ficheux2021} for $CZ_{|11\rangle \leftrightarrow |21\rangle}$. In this scheme, two identical microwave
drives are applied to the two fluxonium qubits, as described by Eq.~(\ref{eq27}), and the drive frequency is placed between
the two nearby plasmon transitions, namely $|11\rangle \leftrightarrow |21\rangle$ and $|10\rangle \leftrightarrow |20\rangle$.
More explicitly, following Ref.~\cite{Ficheux2021}, the drive detuning $\delta$ with respect to the frequency of
the $|11\rangle \leftrightarrow |21\rangle$ transition is given by $\delta = \eta\delta\omega_{12,i}$, where
\begin{equation}
\begin{aligned}\label{eq28}
\eta=\frac{r^2-\sqrt{(r^2-1)^2+r^2}}{r^2-1},r=\frac{|\langle 11|\hat n_{1}+\hat n_{2}|21\rangle|}{|\langle 10|\hat n_{1}+\hat n_{2}|20\rangle|},
\end{aligned}
\end{equation}
the drive amplitude is
\begin{equation}
\begin{aligned}\label{eq29}
&\frac{A}{\delta\omega_{12,i}}=\frac{\sqrt{1-\eta^{2}}}{|\langle 10|\hat n_{1}+\hat n_{2}|20\rangle|},
\end{aligned}
\end{equation}
and the gate length is approximately $2\pi/\delta\omega_{12,i}$.

Note that the analytical estimations of the gate parameters are based on square pulses. Here, we consider using
a flat-top pulse with cosine ramps, and the length of the flat-top part is fixed at $2\pi / \delta\omega_{12,i} \approx 90.6\,\rm ns$.
For each ramp time, based on the analytical estimations, the drive frequency and amplitude are further optimized by minimizing
leakage and phase errors. Accordingly, the red dots in Fig.~\ref{fig11}(a) show the CZ gate error as a function of
ramp time. One can see that although the gate performance is indeed improved compared to a gate of the same length using the
DRAG scheme (see the inset), the lowest gate error remains nearly an order of magnitude above $10^{-4}$. Examining the system
dynamics reveals that the gate performance is currently limited by the spurious off-resonant transition, such as
$|01\rangle \leftrightarrow |02\rangle$, on the neighboring qubit $Q_2$. This is to some extent expected, since the
plasmon frequency detuning ($\Delta_{p}=\omega_{p,2}-\omega_{p,1}$) for the two fluxonium qubits in the current system is approximately
$90\,\rm MHz$, which is far smaller than that in Ref.~\cite{Ficheux2021}, where the detuning is around $300\,\rm MHz$.

To resolve the leakage issue related to off-resonant driving of the neighboring qubit's plasmon transition, we consider
combining the DRAG scheme with the synchronized scheme. More specifically, based on the synchronized pulse design, a DRAG
correction is added to suppress leakage due to the off-resonant transition $|01\rangle \leftrightarrow |02\rangle$ (see Appendix~\ref{App_A}).
Moreover, as shown in Fig.~\ref{fig10}(b), since $\delta\omega_{i,12}$ is only about $6\,\rm MHz$, the DRAG correction is
also expected to suppress leakage from the off-resonant transition $|11\rangle \leftrightarrow |12\rangle$. To examine
this combined approach, we fix the total gate length at $100\,\rm ns$ and optimize the drive parameters similarly
as before.

As illustrated by the orange dots in Fig.~\ref{fig11}(a), which plot the gate error as a function of ramp time, this combination
can suppress the gate error below $10^{-5}$. In addition, Fig.~\ref{fig11}(b) further shows the gate error versus gate length
for varying ramp times. These results suggest that by combining the DRAG scheme with the synchronized scheme, sub-100-ns CZ gates
with intrinsic gate error below $10^{-4}$ can indeed be achieved in the F-LTC-F system, even with a state-dependent frequency
shift of approximately $10\,\rm MHz$.

\subsection{Sub-100-ns CZ gate realization: Required coupling strength}\label{SecIIIC}

\begin{figure}[tbp]
\begin{center}
\includegraphics[keepaspectratio=true,width=\columnwidth]{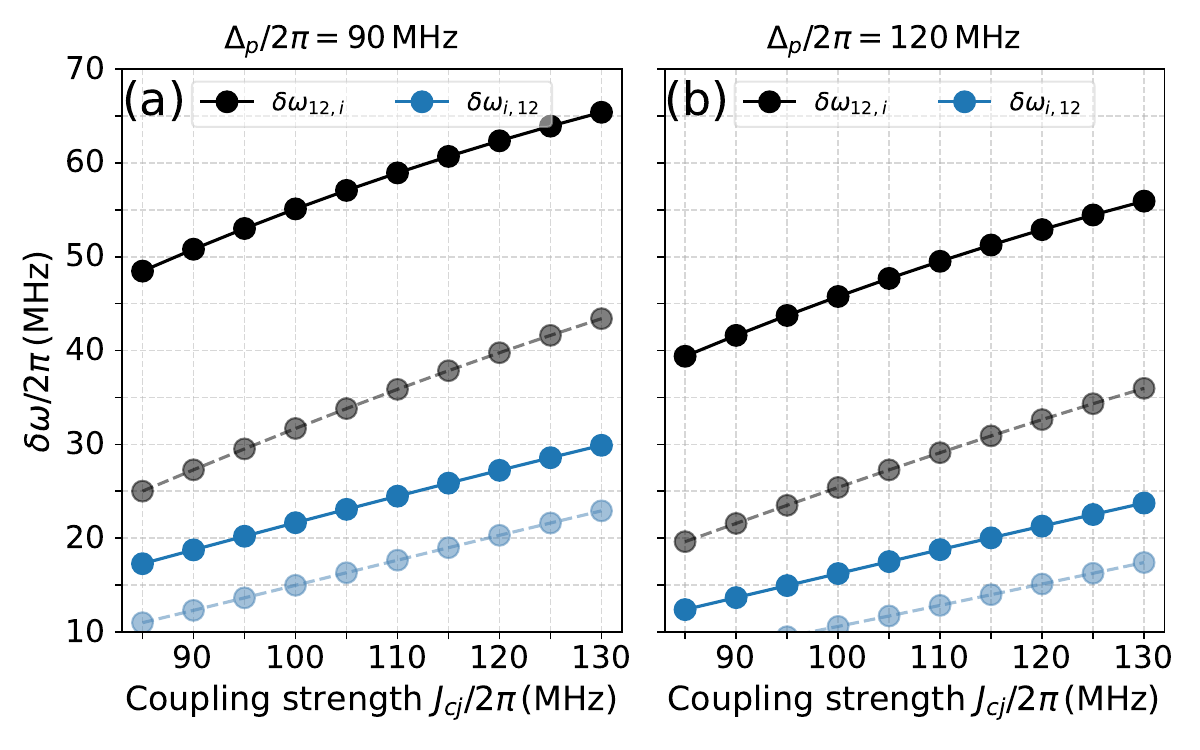}
\end{center}
\caption{State-dependent frequency shifts at different coupling strengths for the F-LTC-F system with a coupler length
of $1\,\rm cm$. (a) State-dependent frequency
shifts as a function of $J_{cj}$ for two different plasmon frequencies (i.e., different plasmon-coupler detunings).
Here, the solid and dashed lines denote the results for $Q_{1}$ plasmon frequencies of $5.367\,\rm GHz$
and $5.450\,\rm GHz$, respectively, while the plasmon frequency of $Q_{2}$ is $90\,\rm MHz$ above
that of $Q_{1}$. (b) Same as (a), but for a plasmon detuning of $120\,\rm MHz$.}
\label{fig12}
\end{figure}

Having discussed CZ gate realization in two specific F-LTC-F systems in the preceding subsections, we
now offer several comments regarding the following question: within the proposed F-LTC-F coupling
architecture, what level of coupling strength, in particular, the fluxonium-LTC coupling
strength $J_{cj}$, is necessary to achieve sub-100-ns CZ gates? Here, we focus on $J_{cj}$ because
strong fluxonium interactions and consequently fast gate speeds require a large $J_{cj}$, which in
turn necessitates a large coupling capacitor. Such a capacitor can introduce significant coupling loading
issues for fluxonium qubits (particularly in qubit lattices with high-degree connectivity, as mentioned
before and will be further discussed in the following section), given that their charge energy is constrained
to approximately $1\,\rm GHz$. As mentioned before, the issue is further exacerbated by the fact that the effective
capacitance of the CPW resonator far exceeds that of couplers based on lumped elements, such as the transmon coupler.
Given this constraint, a design must balance fast gate speeds against a preference for lower $J_{cj}$.

Given the preceding analysis, we focus here on the microwave-activated CZ gate ($CZ_{|11\rangle \leftrightarrow |21\rangle}$) realization within F-LTC-F systems.
Following the results presented in Figs.~\ref{fig7} and~\ref{fig11}, as well as those discussed in
Ref.~\cite{Ficheux2021}, when targeting sub-100 ns gates with intrinsic errors below $10^{-4}$, the
state-dependent frequency shift $\delta\omega$ should exceed $10\,\rm MHz$. Generally, the magnitude of
such shifts, which arises from the energy level repulsion induced by plasmon interactions
$|12\rangle\leftrightarrow|21\rangle$ (see Fig.~\ref{fig8}), can be approximated as~\cite{Chow2013}
\begin{equation}
\begin{aligned}\label{eq30}
&|\delta\omega|\approx\sqrt{\left(\frac{\Delta_{p}}{2}\right)^2+g_{p}^2}-\frac{\Delta_{p}}{2},
\end{aligned}
\end{equation}
where $\Delta_{p}=\omega_{p,2}-\omega_{p,1}$ denotes the detuning of the plasmon transitions of the two
fluxoniums (with $\omega_{p,2} > \omega_{p,1}$), and $g_{p}$ represents the plasmon interaction
strength [see Eq.~(\ref{eq26})].

Considering the LTC parameters listed in Table~\ref{tab:fluxonium_parameters} (with a coupler length
of $1\,\rm cm$ and a fixed inter-resonator inductive coupling) and $L_{S} = 0.5\,\rm nH$, Figure~\ref{fig12}(a) shows
these shifts as a function of the fluxonium-LTC coupling strength $J_{cj}$ for two different plasmon frequencies
of $Q_{1}$, with a detuning of $\Delta_{p}/2\pi = 90\,\rm MHz$ (i.e., the plasmon frequency of $Q_{2}$ is $90\,\rm MHz$ above
that of $Q_{1}$). In addition, Figure~\ref{fig12}(a) shows the same results for a detuning of $\Delta_{p}/2\pi = 120\,\rm MHz$.
The full fluxonium parameters are listed in Appendix~\ref{App_B} (see Table~\ref{tab:fluxonium_parameters_90} and
Table~\ref{tab:fluxonium_parameters_120}).

As expected [see Eqs.~(\ref{eq30}) and~(\ref{eq26})], these results demonstrate that to achieve a large $\delta\omega$ (i.e., fast gates) while
lowering $J_{cj}$ as much as possible, one should favor decreasing $\Delta_{p}$, as well as the detuning between
the resonator mode and the plasmon transition $\Delta_{p,j}$. However, smaller plasmon or resonator-plasmon detunings
can cause strong state hybridization, which could be detrimental, particularly when residual coupling exists~\cite{Zhao2025}.
This concern is especially relevant here because the coupling between high-energy states of the F-LTC-F system generally
lies in the strongly non-dispersive regime. To this end, the LTC featuring an intrinsic zero-point condition could
significantly mitigate these issues. Overall, along with the results shown in Figs.~\ref{fig7} and~\ref{fig11}, the above
analysis shows that for a one-centimeter LTC and the plasmon detuning of $\sim 100\,\rm MHz$, the fluxonium-LTC coupling
could be as low as $80\,\rm MHz$ while still achieving sub-100 ns CZ gates with errors below $10^{-4}$. These observations
suggest that the coupling loading issues within such an F-LTC-F architecture could be largely addressed while maintaining
fast, high-fidelity gates.

We note that even with a fluxonium-LTC coupling strength as high as $125\,\rm MHz$ (e.g., in the system featuring an extended
coupler length of $1.5\,\rm cm$, as discussed in the preceding subsection), coupling loading issues may still be
mitigated through careful design of the coupling layout for such an architecture, as will be elaborated
in the following section. Moreover, increasing the inter-resonator inductive coupling strength (hence $g_{p}$) by employing a
larger coupling junction energy [see Fig~\ref{fig3}(a)] or a terminating inductor [see Figs~\ref{fig2}(e) and~\ref{fig3}(b)] could
also serve to mitigate the coupling loading issues.

\section{Extensions and outlook}\label{SecIV}

In this section, based on specific designs of the coupling layout for the F-LTC-F architecture, we
discuss the challenges and opportunities of integrating the LTC into fluxonium qubit lattices with
degree-four connectivity within a modular architecture. We also assess the feasibility of achieving
even higher-degree connectivity required for implementing complex quantum error correction codes.
Furthermore, we outline an approach for extending the coupler length and enhancing connectivity
by combining the proposed LTC with the multimode cavity QED architecture demonstrated in Ref.~\cite{McKay2015}.
Throughout these discussions, the coupling loading issue for fluxonium qubits is always taken into account, either
explicitly or implicitly.

\subsection{Extensibility to larger qubit lattices: Challenges and Opportunities}\label{SecIVA}
\begin{figure}[tbp]
\begin{center}
\includegraphics[keepaspectratio=true,width=\columnwidth]{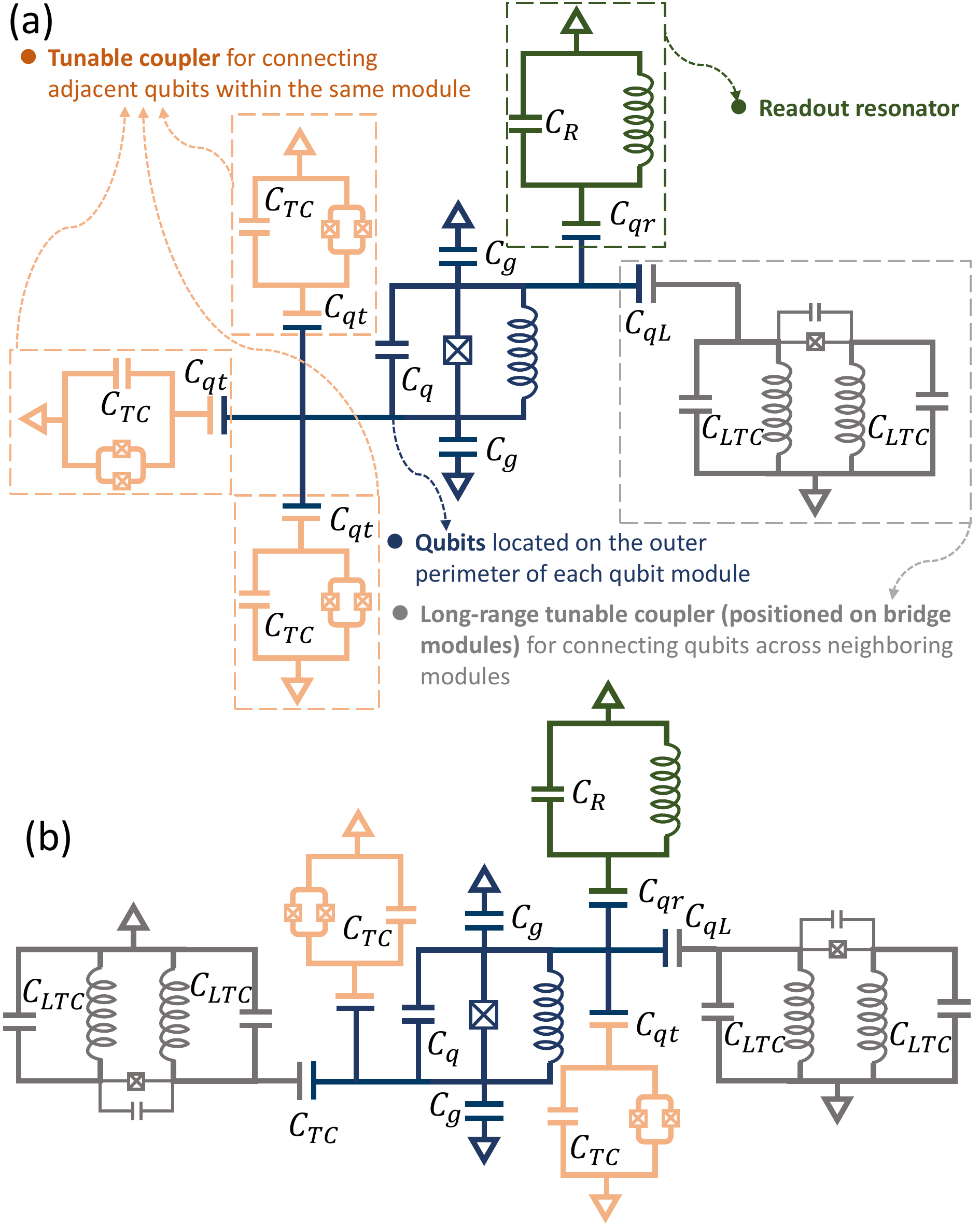}
\end{center}
\caption{Coupling layout for floating fluxonium qubits on the outer perimeter within a modular fluxonium quantum processor featuring
degree-four connectivity. Here, the short-distance TC can be based on single-transmon or double-transmon couplers. Focusing
on the fundamental modes of the CPW resonator within the LTC, the LTC is modeled as two inductively coupled LC oscillators.
In addition, coupling to a readout resonator is also taken into consideration. (a) The '1LTC+3TC' configuration, where
each fluxonium is coupled to three neighbors via short-distance tunable couplers (TCs) on the same module, and to one
neighbor on the adjacent module via the LTC. (b) The '2LTC+2TC' configuration, where each fluxonium is coupled to two
neighbors via short-distance TCs on the same module, and to two neighbors on the adjacent module via the LTC.}
\label{fig13}
\end{figure}

\begin{table}[!htb]
\caption{\label{tab:Capa_parameters} The circuit capacitor parameters for the coupling layout shown in Fig.~\ref{fig14}.}
\begin{ruledtabular}
\begin{tabular}{cccccccc}
Cap. &$C_{TC}$ & $C_{LTC}$ & $C_{R}$ & $C_{q}$ ($C_{g}$) & $ C_{qt}$ & $ C_{qr}$ & $C_{qL}$ \\\hline
(fF) &55.0 & 480.0 & 200.0 & 5.5 (8.0) & 8.0 & 4.0  & 14.5\\
\end{tabular}
\end{ruledtabular}
\end{table}

As shown in Fig.~\ref{fig13}, we consider integrating the LTC (with the length of $1\,\rm cm$) into a modular fluxonium quantum processor with
degree-four connectivity, adopted for the surface code lattice. As mentioned earlier [see Fig.~\ref{fig1}(a)], the
LTC, hosted in the bridge module, is used to connect qubits in nearby qubit modules. In this setting, there exist
two typical coupling configurations for fluxonium qubits on the outer perimeter of each qubit
module, namely '1LTC+3TC' [see Fig.~\ref{fig13}(a)] and '2LTC+2TC' [see Fig.~\ref{fig13}(b)]. Note that, to be practical, coupling
to a readout resonator is also taken into consideration, and a qubit-resonator coupling strength above $50\,\rm MHz$ could be sufficient
for fast, high-fidelity readout, as demonstrated, for example, in Ref.~\cite{Li2025}.

In the '1LTC+3TC' configuration, each fluxonium is coupled to three neighbors via short-distance tunable
couplers (TCs) on the same module, and to one neighbor on the adjacent module via the one-centimeter LTC. The short-distance
tunable couplers can be based on single-transmon or double-transmon couplers, as studied in Ref.~\cite{Zhao2025}, and
we target a fluxonium-TC coupling strength of $300-500\,\rm MHz$ for fast CZ gates~\cite{Rosenfeld2024,Zhao2025}. Given the coupling layout
design for the floating fluxonium qubit shown in Fig.~\ref{fig13}(a), the capacitance matrix of this coupling
circuit is
\begin{equation}\label{eq31}
\mathbf{C_{r}}=\left(
\begin{array}{ccccccc}
\tilde{C}_{1} & -C_q & -C_{qt}& -C_{qt} & -C_{qt} & 0 & 0 \\
-C_q  & \tilde{C}_{2}& 0 & 0 & 0 & -C_{qL} &-C_{qr}\\
-C_{qt} & 0 & \tilde{C}_{s} & 0& 0& 0 &0 \\
-C_{qt} & 0 &0  & \tilde{C}_{s} & 0 &0 & 0\\
-C_{qt} & 0 &0  & 0& \tilde{C}_{s} & 0 &0 \\
0 & -C_{qL} & 0 &0 &0 & \tilde{C}_{l} & 0\\
0 & -C_{qr} & 0 & 0 & 0 & 0 & \tilde{C}_{r} \\
\end{array}
\right)
\end{equation}
with
\begin{equation}
\begin{aligned}\label{eq32}
&\tilde{C}_{1}=C_q+C_g+3C_{qt},
\\&\tilde{C}_{2}=C_q+C_g+C_{qL}+C_{qr},
\\&\tilde{C}_{s}=C_{TC}+C_{qt},
\\&\tilde{C}_{l}=C_{LTC}+C_{qL},
\\&\tilde{C}_{r}=C_{R}+C_{qr}.
\end{aligned}
\end{equation}
Here, as noted in Sec.~\ref{SecIIA}, focusing on the fundamental modes of the CPW resonator within
the LTC, the LTC can be modeled as two inductively coupled LC oscillators, and we have adopted this
approach in this section.

To identify the qubit-relevant modes while removing the free modes (F.M.)~\cite{Kerman2020,Ding2021}, which only contribute charge terms
to the full system Hamiltonian and thus do not participate in the system dynamics, from the circuit (i.e., the floating
fluxonium qubit), we consider the following transformation of the phase variables with respect to
the transformation matrix
\begin{equation}\label{eq33}
\mathbf{S}=\left(
\begin{array}{ccccccc}
1 & 1 & 0& 0 & 0 & 0 & 0 \\
1  & -1& 0 & 0 & 0 & 0 &0 \\
0 & 0 & 1 & 0& 0& 0 &0 \\
0 & 0 &0  & 1 & 0 &0 & 0\\
0 & 0 &0  & 0& 1 & 0 &0 \\
0 & 0 & 0 &0 &0 & 1 & 0\\
0 & 0 & 0 & 0 & 0 & 0 & 1 \\
\end{array}
\right),
\end{equation}
which yields $C = S^{-1} C_{r} S^{-1}$. Accordingly, after removing the free mode, i.e., taking $[\mathbf{C}^{-1}] \equiv \operatorname{Tr}_{\text{F.M.}} C^{-1}$ (tracing out the elements relevant to the free mode), the charge energy for each circuit element and their
coupling strength can be described as follows
\begin{equation}
\begin{aligned}\label{eq34}
&E_{C,k}\equiv \frac{e^{2}[\mathbf{C}^{-1}]_{k,k}}{2}=\frac{e^{2}}{2C_{\Sigma k}},
\\&J_{jk}\equiv 4e^{2}[\mathbf{C}^{-1}]_{j,k} \,(j\neq k).
\end{aligned}
\end{equation}
A similar procedure and the corresponding results also apply to the '2LTC+2TC' coupling configuration, where
each fluxonium is coupled to two neighbors via short-distance TCs on the same module and to two neighbors on the
adjacent module via the LTC, as shown on Fig.~\ref{fig13}(b).

\begin{table}[!htb]
\caption{\label{tab:degree4_parameters} The system Hamiltonian parameters for the coupling layout design with degree-four
connectivity, as shown in Fig.~\ref{fig13}}
\begin{ruledtabular}
\begin{tabular}{ccc}
Energy/$(2\pi)\,$(GHz) & 1LTC+3TC & 2LTC+2TC\\\hline

$E_{C,TC}$ & 0.317 & 0.317\\

$E_{C,LTC}$ & 0.040 & 0.040\\

$E_{C,R}$ & 0.095 & 0.095\\

$E_{C,q}$ & 1.009 & 1.023\\

$ J_{qt}$ & 0.485 & 0.480 (0.559)\\

$ J_{qL}$ & 0.125 & 0.111 (0.129)\\

$J_{qr}$ & 0.083 & 0.086\\
\end{tabular}
\end{ruledtabular}
\end{table}

Given the circuit capacitor parameters summarized in Table~\ref{tab:Capa_parameters} (some of
which are adopted from Ref.~\cite{Ding2023t}, providing capacitor values for an actual layout design
of two fluxonium qubits coupled via a short-range TC), Table~\ref{tab:degree4_parameters} lists the corresponding
circuit Hamiltonian parameters for both configurations according to Eq.~(\ref{eq34}). Comparing the parameters in
Table~\ref{tab:degree4_parameters} with those in Table~\ref{tab:fluxonium_parameters}, one finds that they are
comparable. Thus, one may be confident that the coupling loading issues can be mitigated and that the proposed
LTC design can indeed enable the construction of a modular fluxonium quantum processor.

\begin{figure}[tbp]
\begin{center}
\includegraphics[keepaspectratio=true,width=\columnwidth]{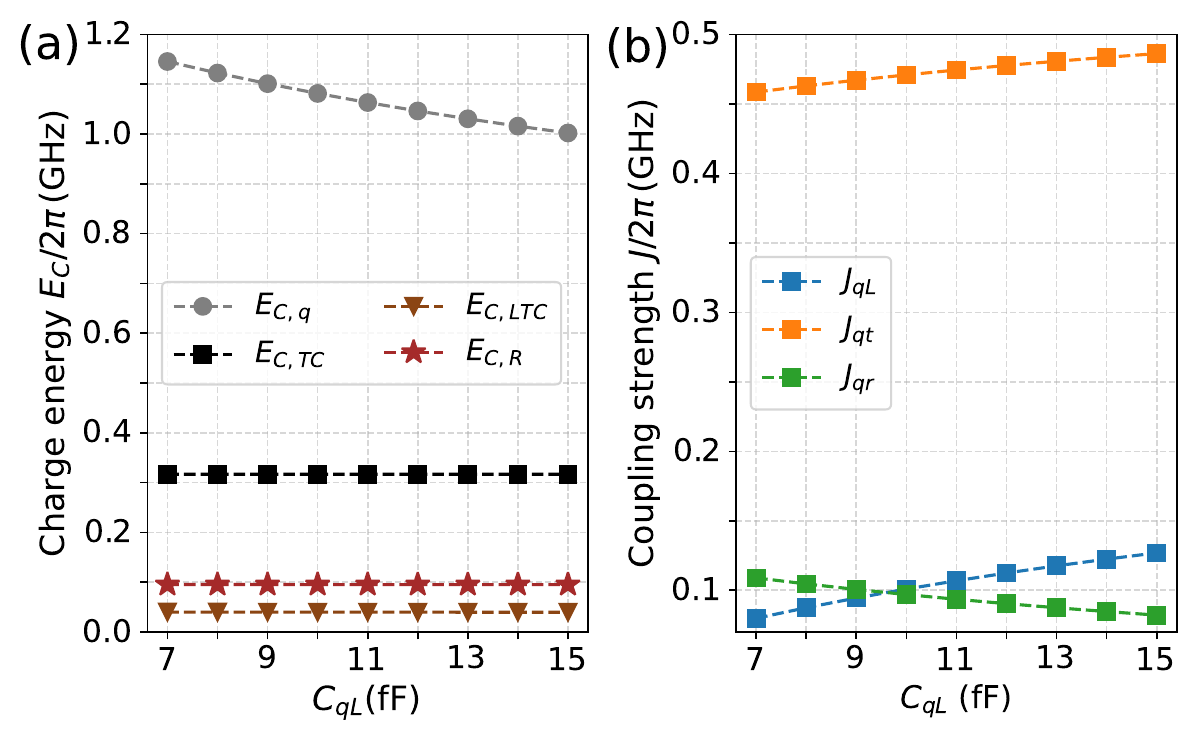}
\end{center}
\caption{System Hamiltonian parameters for the '1LTC+3TC' configuration as a function of the coupling
capacitor $C_{qL}$. The other capacitor parameters are kept the same as in Table~\ref{tab:Capa_parameters}.
(a) Charge energies of the fluxonium qubit, TC, LTC, and readout resonator versus $C_{qL}$. (b) Coupling
strengths among qubits and couplers versus $C_{qL}$.}
\label{fig14}
\end{figure}

Meanwhile, as shown in Table~\ref{tab:Capa_parameters}, the coupling capacitor $C_{qL}$ is $14.5\,\rm fF$ here. While
this value is still realizable, it is significantly larger than those of the other coupling capacitors, making its
practical implementation nontrivial. As mentioned earlier, one may therefore seek to lower this capacitance while
still achieving sub-100 ns gates with an intrinsic error below $10^{-4}$. Accordingly, Figure~\ref{fig14} shows
the circuit Hamiltonian parameters as a function of $C_{qL}$ for the '1LTC+3TC' configuration, with the other
capacitor parameters kept the same as in Table~\ref{tab:Capa_parameters}. It can be seen that $C_{qL}$ can be
reduced to as low as $8\,\rm fF$, matching the short-range TC coupling capacitor $C_{qt}$, while still yielding
a coupling strength of $87.45\,\rm MHz$, which is sufficient for sub-100 ns gates with an error
below $10^{-4}$ (see Fig.~\ref{fig12}).

\begin{figure}[tbp]
\begin{center}
\includegraphics[keepaspectratio=true,width=\columnwidth]{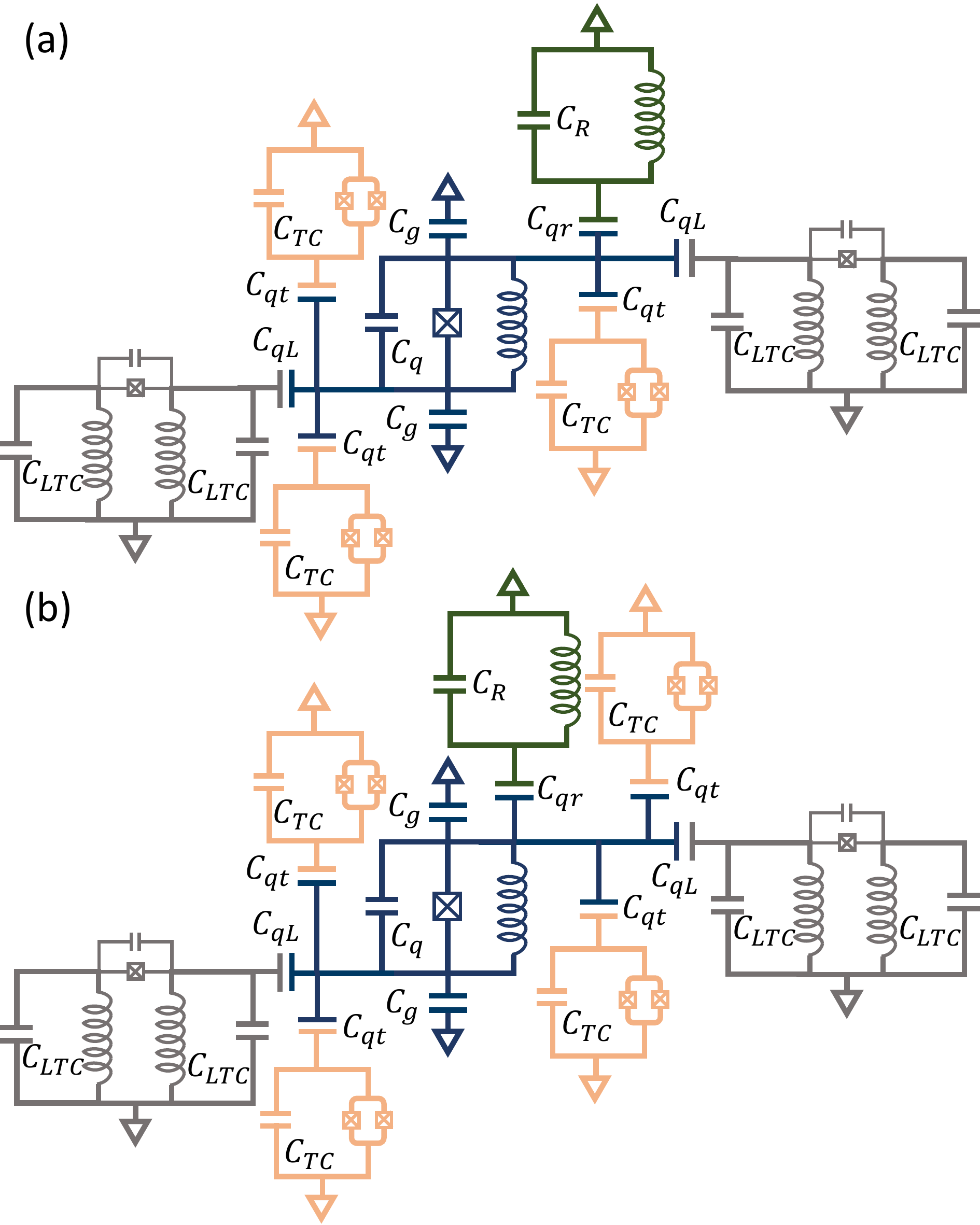}
\end{center}
\caption{Coupling layout for floating fluxonium qubits for fluxonium quantum processor featuring
higher connectivity needed for more complex quantum error correction codes. (a) The '2LTC+3TC' configuration, where
each fluxonium is coupled to three neighbors via short-distance tunable couplers (TCs) and to two
non-neighbors via the LTC. (b) The '2LTC+4TC' configuration, where each fluxonium is coupled to four
near neighbors via short-distance TCs and to two non-neighbors via the LTC.}
\label{fig15}
\end{figure}

Before concluding this subsection, and in light of recent efforts toward implementing quantum
low-density parity-check (qLDPC) codes, which require high connectivity and long-range
coupling~\cite{Breuckmann2021}, we examine the feasibility of integrating the F-LTC-F architecture
for such purposes. Figures~\ref{fig15}(a) and~\ref{fig15}(b) show two typical coupling networks, i.e., the '2LTC+3TC' configuration
and the '2LTC+4TC' configuration, required for the qLDPC codes proposed in Refs.~\cite{Bravyi2024,Shaw2025}.
Following the above analysis and considering the circuit parameters summarized in
Table~\ref{tab:Capa_parameters}, Table~\ref{tab:degree6_parameters} lists the corresponding circuit Hamiltonian parameters
for the two configurations. These circuit parameter analyses, based on feasible capacitor values, suggest that
the LTC can achieve sufficient coupling strength while mitigating coupling loading issues. Therefore, we expect
that the LTC could provide a practical pathway toward building fluxonium quantum processors capable of
supporting qLDPC codes.

\begin{table}[!htb]
\caption{\label{tab:degree6_parameters} The system Hamiltonian parameters for the coupling layout design with higher degree connectivity, as shown in Fig.~\ref{fig15}}
\begin{ruledtabular}
\begin{tabular}{ccc}
Energy$/(2\pi)\,$(GHz) & 2LTC+3TC & 2LTC+4TC \\\hline

$E_{C,TC}$ & 0.315 (0.316) & 0.315 (0.314) \\

$E_{C,LTC}$ & 0.040 & 0.040 \\

$E_{C,R}$ & 0.095 & 0.095 \\

$E_{C,q}$ & 0.852 & 0.792 \\

$ J_{qt}$ & 0.452 (0.414) & 0.423 (0.382) \\

$ J_{qL}$ & 0.104 (0.096) & 0.098 (0.088) \\

$J_{qr}$ & 0.070 & 0.059 \\
\end{tabular}
\end{ruledtabular}
\end{table}

\subsection{Toward even longer couplers and higher connectivity}\label{SecIVB}

\begin{figure}[tbp]
\begin{center}
\includegraphics[keepaspectratio=true,width=\columnwidth]{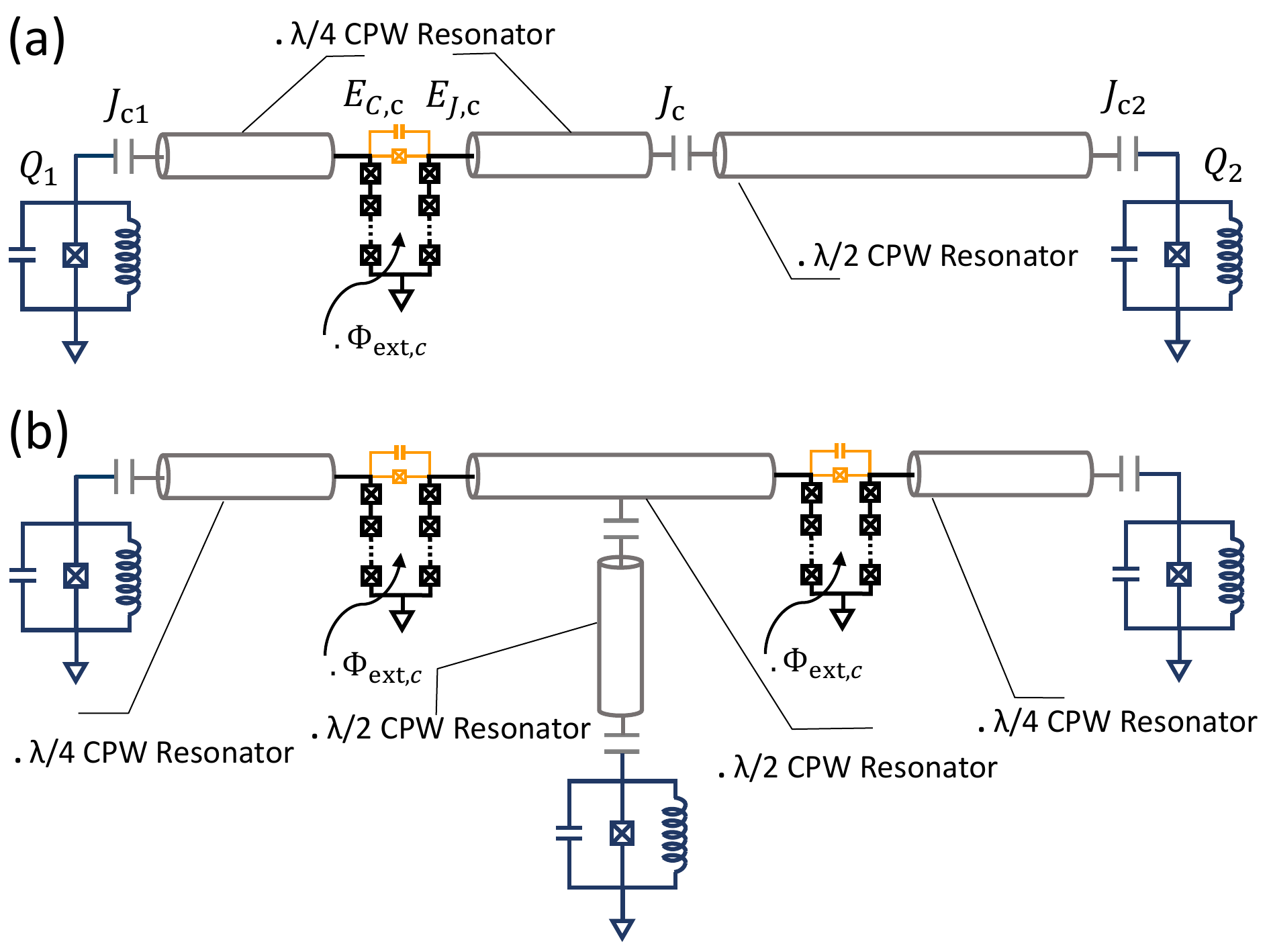}
\end{center}
\caption{Long-range tunable coupler design that embeds the LTC within a multimode coupler to achieve even longer range
distances and enable high connectivity. (a) A three-mode long-range tunable coupler consisting of an LTC capacitively
coupled to a $\lambda/2$ CPW resonator. Focusing on the fundamental modes of these CPW resonators, the coupler
can be modeled as a linear chain of three coupled oscillators with alternating tunable and fixed interactions.
(b) The three-mode long-range tunable coupler design for increasing connectivity, in which each fluxonium qubit is
coupled to other qubits via three CPW resonators, with one of the resonators shared among all three qubits.}
\label{fig16}
\end{figure}

\begin{table}[!htb]
\caption{\label{tab:three_mode_parameters} The circuit Hamiltonian parameters of the coupled fluxonium system
shown in Fig.~\ref{fig16}(a).}
\begin{ruledtabular}
\begin{tabular}{cccc}
(GHz)&
$E_C/2\pi$ &
$E_L/2\pi$ &
$E_J/2\pi$ \\\hline
Fluxonium $Q_{1}$ & 1.00 & 0.80 & 6.85 \\
Fluxonium $Q_{2}$ & 1.00 & 0.75 & 6.90  \\
$\lambda/4$-Resonator & 0.0403 & 74.7550 & $-$ \\
$\lambda/2$-Resonator & 0.0202 & 149.5100 & $-$ \\
\hline
\hline
(GHz) & $J_{c1(c2)}/2\pi$  & $J_{c}/2\pi$  & $E_{J,c}/2\pi$  \\\hline
Coupling strengths & 0.125 (0.065) & 0.030  & 99.336
\end{tabular}
\end{ruledtabular}
\end{table}

The discussions in Sec.~\ref{SecIIIB} and Sec.~\ref{SecIVA} suggest that achieving even longer couplers
and higher connectivity solely with the proposed LTC design is a nontrivial task. Here, we therefore
consider combining the proposed LTC design with the multimode coupler design demonstrated in
Ref.~\cite{McKay2015} to achieve these goals.

As shown in Fig.~\ref{fig16}(a), we first consider a three-mode long-range tunable coupler consisting of an
LTC capacitively coupled to a $\lambda/2$ CPW resonator. As in Sec.~\ref{SecIIA}, when focusing on the fundamental
modes of these CPW resonators, the coupler can be described by the following Hamiltonian
\begin{equation}
\begin{aligned}\label{eq35}
\hat{H}_{\rm MLC}= &\hat{H}_{\rm LTC} +\left[4 E_{C,c} \hat{n}_{c}^2 + \frac{E_{L,c}}{2}\hat\varphi_{c}^2\right]+J_{c} \hat{n}_{c2} \hat{n}_{c},
\end{aligned}
\end{equation}
where the LTC Hamiltonian $\hat{H}_{\rm LTC}$ is given in Eq.~(\ref{eq10}), the subscript $c$ denotes the $\lambda/2$ CPW
resonator, and $J_{c}$ denote the strength of the capacitive coupling between the LTC and the $\lambda/2$ CPW resonator.

\begin{figure}[tbp]
\begin{center}
\includegraphics[keepaspectratio=true,width=\columnwidth]{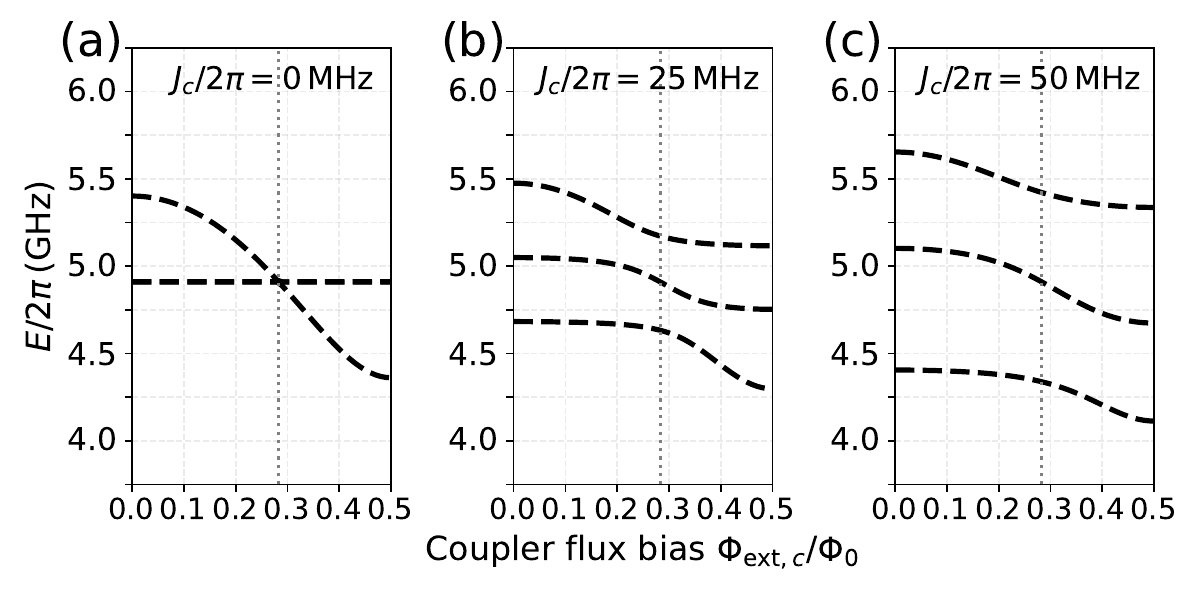}
\end{center}
\caption{Eigenmode frequencies of the three-mode long-range tunable coupler (the eigenvalues of the coupler
Hamiltonian) versus the coupler bias applied to the LTC part for three different capacitive coupling strengths
$J_{c}$, i.e., (a) $0\,\rm MHz$, (b) $25\,\rm MHz$, and (c) $50\,\rm MHz$. The vertical dashed line marks the
bias at which the LTC's inter-resonator inductive coupling is turned off.}
\label{fig17}
\end{figure}

The above Hamiltonian corresponds to a linear chain of three coupled oscillators with alternating tunable
and fixed interactions. Using the CPW parameters listed in Table~\ref{tab:CPW_parameters}, we consider a
specific design based on a one-centimeter LTC capacitively coupled to a one-centimeter $\lambda/2$ CPW
resonator. The three fundamental modes of the three CPW resonators are degenerate, with the associated charge and
inductive energies listed in Table~\ref{tab:three_mode_parameters}. Accordingly, Figure~\ref{fig17} shows the three
eigenmodes of the coupler as a function of the coupled flux bias applied to the LTC part, for three different capacitive
coupling strength $J_{c}$. When $J_{c}=0$, the three modes are degenerate at the bias point $0.283$, again agreeing
well with the analytical result in Eq.~(\ref{eq22}), even with an additional $\lambda/2$ CPW resonator
present. Meanwhile, as $J_{c}$ increases, the degeneracy breaks down as expected.

Note that for three degenerate modes (with frequency $\omega_c$) coupled with the same strength $g$, the
three eigenmode frequencies should be approximately $\omega_c - \sqrt{2}g$, $\omega_c$, and $\omega_c + \sqrt{2}g$~\cite{McKay2015}.
However, this approximation breaks down when the non-rotating coupling terms become non-negligible, i.e., when the
RWA is no longer valid, as is the case in our work (recalling that the maximum inductive inter-resonator coupling strength
can reach $\sim250\,\rm MHz$).

\begin{figure}[tbp]
\begin{center}
\includegraphics[keepaspectratio=true,width=\columnwidth]{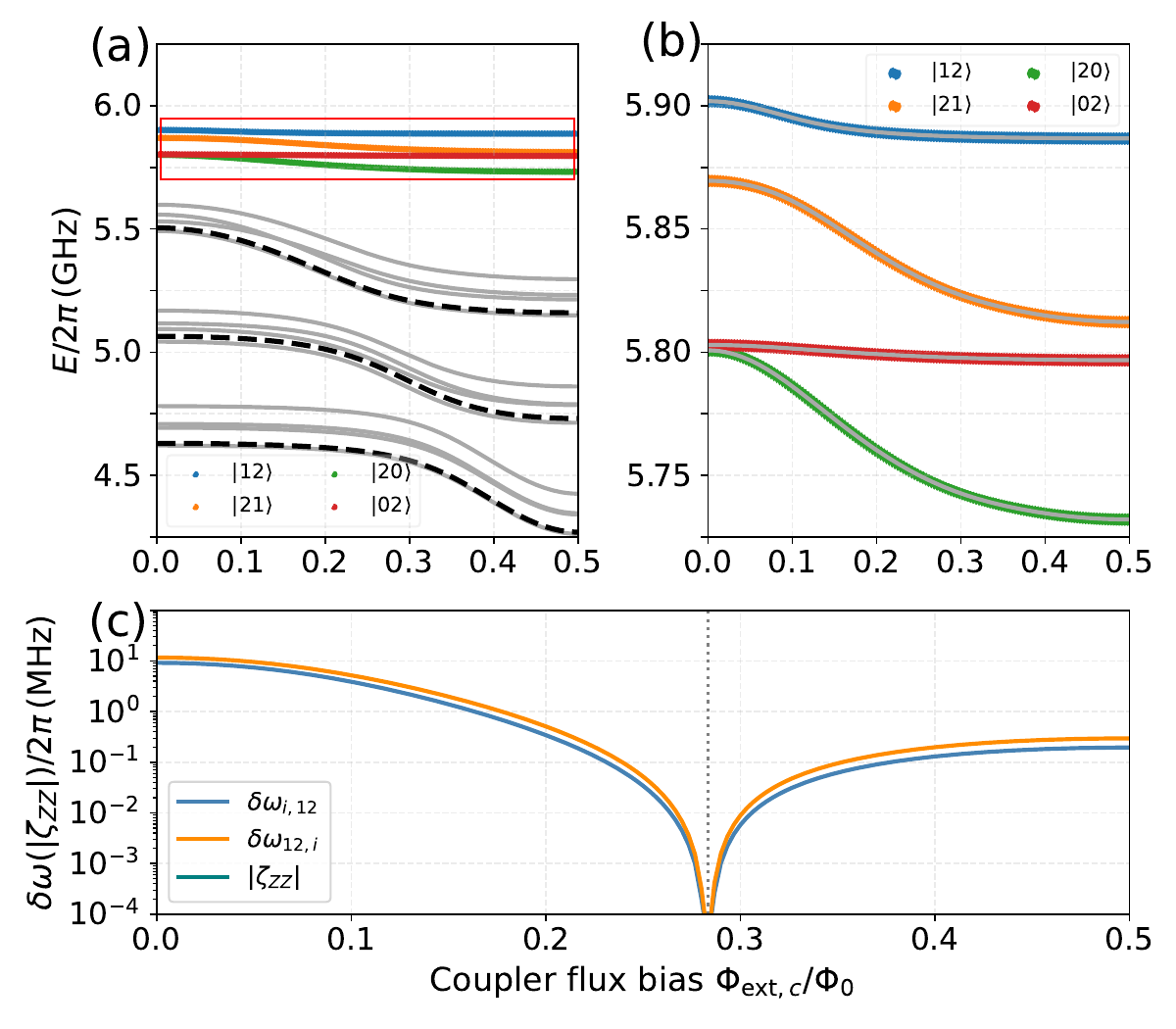}
\end{center}
\caption{System spectrum and coupler-mediated plasmon interactions of fluxonium qubits coupled
via the three-mode coupler. (a) Non-computational energy levels (for the $|1\rangle \leftrightarrow |2\rangle$ plasmon transition)
and eigenmode frequencies of the coupler alone (black dashed lines, as in Fig.~\ref{fig17}) versus coupler flux bias
applied to the LTC part. The region highlighted by the red rectangle is enlarged in Fig.~\ref{fig18}(b).
(c) State-dependent frequency shifts for the $|1\rangle \leftrightarrow |2\rangle$ plasmon transition versus
coupler flux bias. The vertical dashed line marks the shift minimum. The ZZ coupling $\zeta_{ZZ}$ remains
below $0.1\,\rm kHz$ across the entire bias range.}
\label{fig18}
\end{figure}

Having provided the analysis of the coupler alone, we now consider two fluxonium qubits coupled through this coupler
design. The full coupled system Hamiltonian is given as follows:
\begin{equation}
\begin{aligned}\label{eq1}
\hat{H}= &\sum_{j=1,2} [4 E_{C,j} \hat{n}^2_j+\frac{1}{2}E_{L,j}(\hat\varphi_j - \varphi_{\text{ext},j})^2-E_{J,j}\cos\hat\varphi_j]
\\& + J_{c1} \hat{n}_1 \hat{n}_{c1} + J_{c2} \hat{n}_2 \hat{n}_{c} +\hat{H}_{\rm MLC}.
\end{aligned}
\end{equation}
Using the fluxonium qubit parameters and the coupling parameters listed in Table~\ref{tab:three_mode_parameters},
Figure~\ref{fig18}(a) shows the high-energy levels of the full coupled system. The region highlighted by the red
rectangle is enlarged in Fig.~\ref{fig18}(b), which displays the high-energy eigenstates associated with the
fluxonium's plasmon transition $|1\rangle \leftrightarrow |2\rangle$. The corresponding state-dependent frequency
shift is plotted in Fig.~\ref{fig18}(c), with a maximum shift of $11.65\,\rm MHz$, suggesting that high-fidelity,
sub-100 ns gates can also be achieved with this coupler design. Moreover, the shift is minimized
below $0.1\,\rm kHz$ at the bias $0.283$, again consistent with our previous analytical analysis, and
the ZZ coupling remains below $0.1\,\rm kHz$ across the entire bias range.

The above analysis shows that the proposed LTC design can be combined with the multimode coupler
design. Such an approach can enable even longer-range tunable couplers for facilitating fast gates
while suppressing quantum crosstalk. Moreover, as shown in Fig.~\ref{fig16}(b), this coupler design
can also be extended to achieve higher qubit connectivity while mitigating the coupling loading
issue for fluxonium qubits. Overall, this combination opens up new directions for designing
long-range tunable couplers for constructing modular quantum processors and implementing
complex quantum error correction codes.

\section{Discussion and conclusion}\label{SecV}

In conclusion, we have presented a comprehensive theoretical analysis of a long-range tunable
coupler (LTC) architecture that addresses a critical bottleneck in the scalability of fluxonium-based
quantum processors. By leveraging CPW resonators interconnected via tunable inductive couplers, this
design circumvents the restricted length, limited connectivity, and parasitic quantum crosstalk that
often accompany existing monolithic, local-coupling schemes. Our analysis quantitatively demonstrates
that this LTC framework enables high-fidelity CZ gates with durations below $100\,\rm ns$ for qubit
separations exceeding one centimeter, achieving intrinsic gate errors below $10^{-4}$ under realistic
circuit parameters. Furthermore, we have shown that this coupling mechanism is readily extensible to
modular lattice configurations, supporting the degree-four connectivity often required for surface
code implementations, while also providing a viable route toward the higher-degree, nonlocal interactions
essential for the realization of quantum low-density parity-check (qLDPC) codes. The explicit consideration
of qubit coupling loading throughout our analysis confirms the practical feasibility of integrating such
interconnects without compromising the characteristic properties that make fluxonium qubits a compelling
platform.

While here we focus on microwave-activated CZ gates based on the engineered interaction for the
plasmon transition $|1\rangle \leftrightarrow |2\rangle$, this method should also be applicable
to other plasmon transitions, such as $|0\rangle \leftrightarrow |3\rangle$, or even to engineering
fluxonium interactions between different plasmon states, such as $|02\rangle \leftrightarrow |31\rangle$.
In addition, besides the microwave-activated scheme, other schemes, such as the parametric drive
scheme~\cite{Moskalenko2021,Zhao2025c}, could also be considered. These explorations could provide greater
flexibility for engineering long-range couplings and achieving fast, high-fidelity gates for widely
separated qubits.

Beyond the immediate implications for the architecture, the coupling mechanisms described in this
work carry broader relevance for superconducting circuits, as the underlying physics could be directly
applicable to long-range tunable couplers based on different coupling designs, as in Refs.~\cite{Peropadre2013,Wulschner2016}, as
well as to alternative qubit modalities, including the transmon qubit. A particularly promising avenue for future exploration
involves the integration of the proposed LTC with the multimode coupler design, an approach we suggest could
further enhance both the interconnect range and the qubit connectivity while preserving low quantum crosstalk.
While this study establishes a robust theoretical foundation for modular integration, subsequent experimental
validation will be essential to characterize the interplay between the detailed design of the long-range coupler and a high-performance
multi-qubit system. Nevertheless, the work presented here provides a clear and actionable pathway for transitioning from
the demonstrated performance of individual fluxonium devices toward the modular, large-scale quantum
processors adapted to diverse quantum error correction codes.

\begin{acknowledgments}
Peng Zhao would like to thank Zhuang Ma, Haonan Xiong, Xiao Liang, and Dong Lan for insightful discussions. The work is supported by the
National Natural Science Foundation of China (Grants No.12204050, No.92576110, and No.12275090). Peng Xu is supported by the National Natural Science
Foundation of China (Grants No.12105146 and No.12175104) and the Program of State Key Laboratory of Quantum Optics
Technologies and Devices (No.KF202505).
\end{acknowledgments}

\appendix

\section{Effective system Hamiltonian}\label{App_A}

Here, for easy reference and to set the notation, we provide the derivation of the effective
Hamiltonian given in Eq.~(\ref{eq25}), following the procedure in Ref.~\cite{Zhao2025}. As
mentioned in Sec.~\ref{SecIIB1}, the inter-resonator inductive coupling is the dominant
coupling within the LTC, thus, omitting the inter-resonator capacitive coupling for the
LTC gives rise to the coupler Hamiltonian
\begin{equation}
\begin{aligned}\label{eqA1}
\hat{H}_{LTC}=&\sum_{j=1,2}[\omega_{c,j}\hat a_{cj}^{\dag}\hat a_{cj}]+g_{c}(\hat a_{c1}+\hat a_{c1}^{\dag})(\hat a_{c2}+\hat a_{c2}^{\dag})
\end{aligned}
\end{equation}
with $g_{c}=g_{c,induc}$ and $\omega_{c,j}=\omega_{c}$ (considering degenerate modes).

The coupler Hamiltonian
can be diagonalized by employing the Bogoliubov transformation
$[\hat a_{c1},\hat a_{c1}^{\dagger},\hat a_{c2},\hat a_{c2}^{\dagger}]^{T}=M [\hat a_{-},\hat a_{-}^{\dagger},\hat a_{+},\hat a_{+}^{\dagger}]^{T}$.
Keeping terms to order $g_{c}/\omega_{c}$, the transformation matrix $M$ is expressed as~\cite{Zhang2023}
\begin{equation}\label{eqA2}
M\approx\left[
\begin{array}{cccc}
\frac{1}{\sqrt{2}} & \frac{g_{c}}{2\sqrt{2}\omega_{c}} & \frac{1}{\sqrt{2}} & -\frac{g_{c}}{2\sqrt{2}\omega_{c}} \\
\frac{g_{c}}{2\sqrt{2}\omega_{c}} & \frac{1}{\sqrt{2}} & -\frac{g_{c}}{2\sqrt{2}\omega_{c}} & \frac{1}{\sqrt{2}} \\
-\frac{1}{\sqrt{2}} & -\frac{g_{c}}{2\sqrt{2}\omega_{c}} & \frac{1}{\sqrt{2}} & -\frac{g_{c}}{2\sqrt{2}\omega_{c}}\\
-\frac{g_{c}}{2\sqrt{2}\omega_{c}} & -\frac{1}{\sqrt{2}} & -\frac{g_{c}}{2\sqrt{2}\omega_{c}} & \frac{1}{\sqrt{2}}\\
\end{array}
\right].
\end{equation}
and the eigenmode frequencies of the coupler are $\omega_{\pm}=\sqrt{\omega_{c}^{2}\pm 2g_{c}\omega_{c}}\approx\omega_{c}\pm g_{c}$.
Accordingly, inserting Eq.~(\ref{eqA2}) into Eq.~(\ref{eqA1}), the interaction part of the full system
Hamiltonian given in Eq.~(\ref{eq24}) can be rewritten as
\begin{equation}
\begin{aligned}\label{eqA3}
\hat{H}_{p,I}&=\sum_{j=1,2}[g_{p,j}(\hat p_{j}^{\dag}\hat a_{cj}+\hat p_{j}\hat a_{cj}^{\dag})]\\
&=g_{p,1}\left(\frac{1}{\sqrt{2}}-\frac{g_{c}}{2\sqrt{2}\omega_{c}}\right)(\hat a_{-}\hat p_{1}^{\dag}+\hat a_{-}^{\dag}\hat p_{1})\\
&\quad +g_{p,1}\left(\frac{1}{\sqrt{2}}+\frac{g_{c}}{2\sqrt{2}\omega_{c}}\right)(\hat a_{+}\hat p_{1}^{\dag}+\hat a_{+}^{\dag}\hat p_{1})\\
&\quad -g_{p,2}\left(\frac{1}{\sqrt{2}}-\frac{g_{c}}{2\sqrt{2}\omega_{c}}\right)(\hat a_{-}\hat p_{2}^{\dag}+\hat a_{-}^{\dag}\hat p_{2})\\
&\quad +g_{p,2}\left(\frac{1}{\sqrt{2}}+\frac{g_{c}}{2\sqrt{2}\omega_{c}}\right)(\hat a_{+}\hat p_{2}^{\dag}+\hat a_{+}^{\dag}\hat p_{2}).
\end{aligned}
\end{equation}

The above equation describes two fluxonium transitions coupled via two independent modes. Assuming that both
modes are dispersively coupled to the fluxonium transitions, an effective interaction Hamiltonian can be
obtained by adiabatically eliminating the direct fluxonium-mode coupling terms in the above equation~\cite{Bravyi2011}, yielding
\begin{equation}
\begin{aligned}\label{eqA4}
\hat{H}_{p,I}^{({\rm eff})}=g_{p,{\rm eff}}(\hat p_{1}^{\dag}\hat p_{2}+\hat p_{1}\hat p_{2}^{\dag}),
\end{aligned}
\end{equation}
with the effective coupling strength
\begin{equation}
\begin{aligned}\label{eqA5}
g_{p,{\rm eff}}=&\frac{g_{p,1}g_{p,2}}{2}\left(\frac{1}{\sqrt{2}}+\frac{g_{c}}{2\sqrt{2}\omega_{c}}\right)^{2}
\sum_{j=1,2}\left(\frac{1}{\Delta_{p,j+}}\right)\\
&-\frac{g_{p,1}g_{p,2}}{2}\left(\frac{1}{\sqrt{2}}-\frac{g_{c}}{2\sqrt{2}\omega_{c}}\right)^{2}
\sum_{j=1,2}\left(\frac{1}{\Delta_{p,j-}}\right),
\end{aligned}
\end{equation}
where $\Delta_{p,j\pm} = \omega_{p,j} - \omega_{\pm}$ represents the detuning between the fluxonium transition
and the coupler eigenmode. Keeping terms to order $g_{c}/\omega_{c}$, the coupling strength is approximated as
\begin{equation}
\begin{aligned}\label{eqA6}
g_{kl,{\rm eff}}\approx\frac{g_{p,1}g_{p,2}g_{c}}{2}
\sum_{j=1,2}\left(\frac{1}{\Delta_{p,i}^{2}}+\frac{1}{\omega_{c}}\frac{1}{\Delta_{p,j}}\right).
\end{aligned}
\end{equation}
This recovers the formula given in Eq.~(\ref{eq26}).

\section{Microwave-activated CZ gates}\label{App_B}

As mentioned in Sec.~\ref{SecIIIA}, in this work we consider the implementation of microwave-activated CZ gates
based on applying simultaneous microwave drives with identical amplitude and frequency to both fluxonium qubits.
Accordingly, two pulse shape designs are utilized for this purpose, as follows.

(i) \emph{Cosine}, which is defined by
\begin{equation}
\begin{aligned}\label{eqB1}
&A(t)=\Omega_{d}\left(1-\cos\frac{2\pi t}{t_{g}}\right),
\end{aligned}
\end{equation}
where $\Omega_{d}$ denotes the peak drive amplitude and $t_g$ represents the full gate
length (excluding the coupler bias ramping).

(ii) \emph{Flat-top cosine}, which is given by
\begin{align}
A(t)\equiv
\begin{cases}
\Omega_{d}\frac{1-\cos{\left(\pi \frac{t}{t_r}\right)}}{2}  \;, &0<t<t_r\\
\Omega_{d}\;,  &t_r<t<t_g-t_r\\
\Omega_{d}\frac{1-\cos{\left(\pi \frac{t_g-t}{t_r}\right)}}{2} \;, &t_g-t_r<t<t_g
\end{cases}
\label{eqB2}
\end{align}
with the ramp time $t_{r}$.

When seeking to mitigate leading leakage sources from spurious transitions, the DRAG scheme is utilized with
the pulse defined as
\begin{equation}
\begin{aligned}\label{eqB3}
A_{\rm DRAG}(t)=A(t)+i\frac{\alpha}{\delta} \dot{A}(t)
\end{aligned}
\end{equation}
where $\alpha=1$ and $\delta$ denotes the detuning of the spurious transition with respect to the target gate transition.

To tune up the gates, the gate parameters, i.e., the drive peak amplitude $\Omega_{d}$ and the drive frequency $\omega_{d}$, are
optimized by minimizing the leakage and the conditional phase error. To further characterize the intrinsic gate
performance (without the consideration of decoherence), the metric of state-averaged gate fidelity is used.
Specifically, up to single-qubit $Z$ phases, the gate fidelity of the implemented CZ gate is \cite{Pedersen2007}
\begin{equation}
\begin{aligned}\label{eqB4}
F=\frac{{\rm Tr}(\tilde{U}^{\dagger}\tilde{U})+|{\rm Tr}(U_{\rm CZ}^{\dag}\tilde{U})|^{2}}{20},
\end{aligned}
\end{equation}
where $\tilde{U}$ denotes the actual evolution operator given the optimized gate parameters, truncated to the two-qubit
computational subspace spanned by ${|00\rangle,|01\rangle,|10\rangle,|11\rangle}$, and $U_{\rm CZ}$ denotes the ideal CZ gate.

\section{Full fluxonium Hamiltonian parameters}\label{App_C}

\begin{table}[tbp]
\caption{\label{tab:fluxonium_parameters_90}The circuit Hamiltonian parameters of the coupled
fluxonium system shown in Fig.~\ref{fig9}(a) with $\Delta_{p}/2\pi=90\,\rm MHz$. }
\begin{ruledtabular}
\begin{tabular}{cccc}
$\omega_{p1}/2\pi$ (GHz)&$E_C/2\pi$ (GHz) &$E_L/2\pi$ (GHz) & $E_J/2\pi$ (GHz) \\\hline

\multirow{2}{*}{5.367} & 1.00 & 0.80 & 6.55 \\
 & 1.00 & 0.75 & 6.60 \\ \hline
\multirow{2}{*}{5.383} & 1.00 & 0.80 & 6.57 \\
 & 1.00 & 0.75 & 6.62 \\ \hline
\multirow{2}{*}{5.405} & 1.00 & 0.80 & 6.60 \\
 & 1.00 & 0.75 & 6.65 \\ \hline
\multirow{2}{*}{5.411} & 1.00 & 0.80 & 6.607 \\
 & 1.00 & 0.75 & 6.657 \\ \hline
\multirow{2}{*}{5.436} & 1.00 & 0.80 & 6.64 \\
 & 1.00 & 0.75 & 6.69 \\ \hline
\multirow{2}{*}{5.450} & 1.00 & 0.80 & 6.66 \\
 & 1.00 & 0.75 & 6.71 \\ \hline
\multirow{2}{*}{5.458} & 1.00 & 0.80 & 6.67 \\
 & 1.00 & 0.75 & 6.72 \\
\end{tabular}
\end{ruledtabular}
\end{table}

\begin{table}[tbp]
\caption{\label{tab:fluxonium_parameters_120} The circuit Hamiltonian parameters of the coupled fluxonium system
shown in Fig.~\ref{fig9}(b) with $\Delta_{p}/2\pi=120\,\rm MHz$. }
\begin{ruledtabular}
\begin{tabular}{cccc}
$\omega_{p1}/2\pi$ (GHz)&$E_C/2\pi$ (GHz) &$E_L/2\pi$ (GHz) & $E_J/2\pi$ (GHz) \\\hline

\multirow{2}{*}{5.367} & 1.00 & 0.80 & 6.55 \\
 & 1.00 & 0.75 & 6.64 \\ \hline
\multirow{2}{*}{5.390} & 1.00 & 0.80 & 6.58 \\
 & 1.00 & 0.75 & 6.67 \\ \hline
\multirow{2}{*}{5.398} & 1.00 & 0.80 & 6.59 \\
 & 1.00 & 0.75 & 6.68 \\ \hline
\multirow{2}{*}{5.405} & 1.00 & 0.80 & 6.60 \\
 & 1.00 & 0.75 & 6.69 \\ \hline
\multirow{2}{*}{5.420} & 1.00 & 0.80 & 6.62 \\
 & 1.00 & 0.75 & 6.71 \\ \hline
\multirow{2}{*}{5.450} & 1.00 & 0.80 & 6.66 \\
 & 1.00 & 0.75 & 6.75 \\
\end{tabular}
\end{ruledtabular}
\end{table}

For the analysis of the CZ gate performance as a function of the plasmon transition frequency $\omega_{p,1}$ and
detuning $\Delta_{p}$ shown in Fig.~\ref{fig9}, the fluxonium parameters used are summarized in Table~\ref{tab:fluxonium_parameters_90} and Table~\ref{tab:fluxonium_parameters_120}.

\section{Gate errors due to relaxation and dephasing of non-computational gate levels}\label{App_D}

Although the characterized intrinsic CZ gate performance is already very high, the potential actual
implementation entails a number of challenges, the leading one being qubit decoherence. To this
end, we provide a rough estimation of the decoherence effect on the CZ gates, following the discussion
in Refs.~\cite{Ficheux2021,Zhao2025}.

For high-coherence fluxoniums (e.g., with both relaxation and dephasing times reaching milliseconds, as
demonstrated in Ref.~\cite{Somoroff2023}), the leading incoherent gate error could arise from the relaxation
and dephasing of non-computational gate states, i.e., $|02(20)\rangle$ and $|12(21)\rangle$~\cite{Ficheux2021,Ding2023,Abad2025}, and
can be approximated as~\cite{Zhao2025,Abad2025}
\begin{equation}
\begin{aligned}\label{eqD1}
\epsilon=\frac{3}{32}\frac{t_{g}}{T_{1}^{|f\rangle}}+\frac{13}{80}\frac{t_{g}}{T_{\phi,{\rm white}}^{|f\rangle}},
\end{aligned}
\end{equation}
where $T_{1}^{|f\rangle}$ and $T_{\phi,{\rm white}}^{|f\rangle}$ denote the relaxation time and the dephasing (white noise) time of the
non-computational gate level $|f\rangle$.

Assuming typical coherence times of $T_{1}^{|f\rangle}=T_{2}^{|f\rangle}=10\,\mu\rm s$~\cite{Ficheux2021,Ding2023}, we
estimate that for $100\,\rm ns$ CZ gates, the error ($\sim 10^{-3}$) is limited by the decoherence of non-computational
gate states, yielding $\epsilon=0.00175$. Further improvements in the coherence times of these non-computational gate
states are expected to be achievable in the near term, thereby pushing the gate error below $10^{-3}$ or even
approaching $10^{-4}$. Note here that as mentioned in Sec.~\ref{SecIIB1}, in the present system, besides the
decoherence channels of the fluxonium qubit itself, the strong state hybridization due to large fluxonium-coupler
interactions can introduce an additional Purcell-like channel. However, given that the state-of-the-art internal
quality factor of CPW resonators is generally above $10^6$ or even reaches $10^7$, as demonstrated in
Ref.~\cite{Zikiy2025}, such coupler-induced decoherence should not constitute the fundamental limit
for achieving gate errors approaching $10^{-4}$.

\end{document}